\documentclass[aps,prd,twocolumn,nomerge,superscriptaddress,preprintnumbers,superscriptaddress,nofootinbib,floatfix]{revtex4-1}

\usepackage{hyperref}
\usepackage{amsmath,amssymb,amsfonts,amsthm,latexsym,stmaryrd}
\usepackage{marginnote}
\usepackage{color}
\usepackage{braket}
\usepackage{dsfont}

\usepackage{graphicx}

\usepackage{siunitx}
\usepackage{comment}

\def\beq{\begin{equation}}
\def\eeq{\end{equation}}
\newcommand{\bea}{\begin{eqnarray}}
\newcommand{\eea}{\end{eqnarray}}
\def\bfig{\begin{figure}}
\def\efig{\end{figure}}

\def\be{\begin{equation}}
\def\ee{\end{equation}}
\newcommand{\ba}{\begin{eqnarray}}
\newcommand{\ea}{\end{eqnarray}}

\begin{document}

\title{Gravitational wave observations, distance measurement uncertainties, and cosmology}

\author{E.~Chassande-Mottin}
\affiliation{AstroParticule et Cosmologie (APC), Universit\'e Paris Diderot, CNRS/IN2P3, CEA/Irfu, Observatoire de Paris, Sorbonne Paris Cit\'e, 75205 Paris France}
\author{K.~Leyde}
\affiliation{AstroParticule et Cosmologie (APC), Universit\'e Paris Diderot, CNRS/IN2P3, CEA/Irfu, Observatoire de Paris, Sorbonne Paris Cit\'e, 75205 Paris France}
\author{S.~Mastrogiovanni}
\affiliation{AstroParticule et Cosmologie (APC), Universit\'e Paris Diderot, CNRS/IN2P3, CEA/Irfu, Observatoire de Paris, Sorbonne Paris Cit\'e, 75205 Paris France}
\affiliation{Paris Centre for Cosmological Physics}
\author{D.A.~Steer}
\affiliation{AstroParticule et Cosmologie (APC), Universit\'e Paris Diderot, CNRS/IN2P3, CEA/Irfu, Observatoire de Paris, Sorbonne Paris Cit\'e, 75205 Paris France}

\thanks{All authors contributed equally to this work}

\date{\today}

\begin{abstract}
Gravitational waves from the coalescence of compact binaries, together with an associated electromagnetic counterpart, are ideal probes of cosmological models. As demonstrated with GW170817, such multimessenger observations allow one to use the source as a standard siren, the analog of standard candles in conventional astronomy, in order to measure cosmological parameters such as the Hubble constant. No cosmological ladder is needed to estimate the source luminosity distance from the detected gravitational waves. The error on the luminosity distance plays a crucial role in the error budget for the inference of the Hubble constant. In this paper, we provide analytic expressions for the statistical errors on the luminosity distance inferred from gravitational wave data as a function of the sky position and the detector network. In particular, we take into account degeneracy in the parameter space of the gravitational waveform showing that in certain conditions on the gravitational-wave detector network and the source sky position it may not be possible to estimate the luminosity distance of the source. Our analytic approximants shows a good agreement with the uncertainties measured with Bayesian samplers and simulated data. We also present implications for the estimation error on the Hubble constant.
\end{abstract} 

\maketitle

\section{Introduction}

The first direct observations of gravitational waves (GWs) by the LIGO and Virgo collaborations \cite{Abbott:2016blz,LIGOScientific:2018mvr} has opened the possibility of studying astrophysical compact objects and gravity in the strong-field regime (see e.g., \cite{LIGOScientific:2019fpa}). So far, the LIGO and Virgo collaborations have reported the firm detection of eleven compact binary coalescences (CBC) including ten binary black hole mergers (BBH) \cite{LIGOScientific:2018mvr} and one binary neutron star (BNS) merger \cite{TheLIGOScientific:2017qsa} during the first two scientific runs. LIGO and Virgo have recently resumed observations with an improved sensitivity and distance reach, and began to collect new GW events, while future GW detectors such as KAGRA \cite{2012CQGra..29l4007S} and LIGO-India\cite{2013IJMPD..2241010U} are under development.

Besides the intrinsic properties of the GW source itself, it is well known that GWs can be exploited to study cosmology \cite{2012PhRvD..86d3011D}. In particular, as was shown by Schutz \cite{Schutz:1986gp}, CBCs are  cosmological standard  rulers (often referred to as ``standard sirens''), as it is possible to measure their luminosity distance directly from the detected GWs.  Sources in the local universe (redshifts $z \ll 1$) can then be used to determine a new, independent, measurement of  the Hubble constant $H_0$. This type of study is particularly relevant for BNS mergers with electromagnetic counterparts for which the redshift (together with an accurate sky localization) are known. It should be noted that the value of $H_0$ is currently under debate \cite{Jackson2015}: Hubble Space Telescope measurements using supernovae (and Cepheids) measure $H_0 = 73.24\pm1.74\;\SI{}{km\; s^{-1}\;Mpc^{-1}}$ \cite{0004-637X-826-1-56}, whereas cosmic microwave background measurements give $H_0=67.8\pm0.9\;\SI{}{km\; s^{-1}\;Mpc^{-1}}$ \cite{ade2016planck}. 

Other independent determinations of $H_0$ may be crucial to explain this discrepancy. Those include GW-based measurements. Indeed, if the GW measurement of $H_0$ is not consistent with the HST value (both these measurements being at low redshift), this would hint to a bias in the evaluation of the value obtained in \cite{0004-637X-826-1-56}. The opposite case where the GW and HST $H_0$ estimates are compatible opens up a host of interesting ideas such as the possible modification of gravity on large scales \cite{2019LRR....22....1I}. 
So far, the best GW measurement of the Hubble constant is given by the joint electromagnetic and GWs observation of the BNS GW170817, combined with with binary black hole GW  detections and galaxy catalogues. This measurement is $H_0 = 68^{+14.0}_{-7.0} \;\SI{}{km\; s^{-1}\;Mpc^{-1}}$ at $1\sigma$ confidence level \cite{2019arXiv190806060T}. Similar values inferred from GWs observations are also given in \cite{ligo2017gravitational, Abbott:2018wiz, 2019PhRvX...9a1001A}.

%So far, the best GWs measurement of the Hubble constant is given by the joint electromagnetic and GWs observation of the BNS GW170817 \cite{ligo2017gravitational}, which makes use of the identification of the host galaxy, and it is $H_0 = 70.0^{+12.0}_{-8.0} \;\SI{}{km\; s^{-1}\;Mpc^{-1}}$ at $1\sigma$ confidence level \cite{ligo2017gravitational, Abbott:2018wiz, 2019PhRvX...9a1001A}. 

The uncertainty of the $H_0$ estimate is naturally related to the uncertainty on the luminosity distance $D$ of the source, since at low redshift $H_0 \sim cz/ D$. In order to anticipate which error can be expected on $H_0$, it is thus crucial to understand which are the main sources of error in the evaluation of $D$ from GW signals.

Some of the binary parameters have a similar effect on the GW signal, causing degeneracy in their estimation, thus leading to larger measurement uncertainties. The distance $D$ and the inclination $\iota$ of the binary orbital plane enter in the expression of the gravitational wave signal from a binary merger (to leading order) in the combination $(1+\cos^2 \iota)/D$  and $\cos \iota/D$ for the '+' and '$\times$' GW polarizations respectively. 
While $D$ and $\iota$ can be individually estimated from an accurate measurement of the amplitudes of both GW polarizations, they are degenerate when the detector network is essentially sensitive to one polarization \cite{PhysRevD.85.104045}.

In this paper we focus on the distance measurement uncertainty for a given BNS event, and study how it varies as a function of the properties of the source --- including its sky location and orientation --- and of the network of gravitational-wave detectors --- including the number and sensitivity of detectors in the network, and their relative alignment.

Our objective is to provide an {\it analytic} framework in which the following questions can be addressed: (\textit{i}) Are there directions in the sky in which the distance measurement accuracy is significantly better? (\textit{ii}) Do three or more detectors help to break the inclination/distance degeneracy? If yes, by how much does the distance error decrease?

Uncertainties on parameter estimates can be obtained from the Fisher matrix formalism in the limit of large SNR \cite{vallisneri2008use}. This paper revisits and extends the seminal calculations done by Cutler and Flanagan in \cite{cutler1994gravitational} that provides first-order (or ``Gaussian") as well as higher-order (or ``beyond Gaussian") approximations of the estimation errors. 
Those estimates are easily applied to binary mergers at any source sky position, detector network, etc. We obtain skymaps of the predicted errors for present and future detector networks that we compare with the errors obtained with the (computationally demanding) Bayesian sampling algorithms (see e.g., \cite{2015PhRvD..91d2003V}) currently used to perform the astrophysical parameter estimation from the observation data. We find a good agreement except in certain sky positions that we identify and we explain the reason for the discrepancy. We stress the important role played by the degeneracy parameter $\epsilon_d$ defined in Eq.~\eqref{eq:thetadiag}, and draw implications for sources at sky locations where this parameter is close to 1 (likelihood strongly degenerate). Finally, we discuss how our estimates propagate to the errors in the Hubble constant and deduce implications for future observations. 

The paper is organized as follows: in Sec.~\ref{sec:2} we introduce the data analysis background for GW. Sec.~\ref{sec:3} presents the mathematical framework developed to predict the luminosity distance uncertainties.  In Sec.~\ref{sec:4} we discuss several estimates for the luminosity distance uncertainties, and we then compare them to Bayesian analyses of simulated GW signals in LIGO and Virgo noise in Sec.~\ref{sec:5}. Finally in Sec.~\ref{sec:6} we apply the uncertainty estimates to the specific case of GW170817 and to the Hubble constant estimate obtained from this observation.

\section{Data analysis background \label{sec:2}}

In this section we briefly describe the GW signal, focusing on the inspiral phase, and also outline the standard parameter estimation methodology now common in all the GW literature \cite{Jaranowski2012}. In doing so we introduce the relevant notation and expressions used throughout the paper.

\subsection{GW waveforms and detectors}

We assume that the BNS inspirals in quasi-circular orbits, and neglect the effects of both spin and tidal deformability. Furthermore, we assume that the sky position, identified by the right ascension $\alpha$ and declination $\delta$, as well as the redshift $z$ of the source are known thanks to the observation of an electromagnetic counterpart. The GW waveform emitted by the inspiralling BNS in the wave propagation frame \cite{maggiore2008gravitational}, is then given by
\begin{subequations}
\label{eq:waveform_at_source}
\begin{align}
h_+(t) &=\left[\frac{1+v^2}{2D} \right] a_0(t) \cos \varphi_0(t) \\ h_{\times}(t)&=-\left[\frac{v}{D}\right]a_0(t) \sin \varphi_0(t)
\end{align}
\end{subequations}
with
\begin{align}
a_0(t)&=\dfrac{5^{1/4}(G\mathcal{M})^{5/4}}{c^{11/4}} (t_c-t)^{-1/4}\\
\text{and~} \varphi_0(t)&= 2\phi_c+2\:\phi(t-t_c;\mathcal{M})
\end{align}
and where $\phi_c$ is the phase at the coalescence time $t_c$. We consider amplitude and phase evolutions described by the lowest order approximation in the Post-Newtonian expansion given, e.g., in \cite{Sathyaprakash2009}. At this approximation order, the amplitude and phase only depends upon the \textit{chirp mass} $\mathcal{M}= {(m_1 m_2)^{3/5}}{(m_1+m_2)^{-1/5}}$. 

We define the parameter $v\equiv \cos(\iota)$ with $\iota$ the inclination of the orbital plane, i.e., the angle between the angular momentum of the binary and the line-of-sight, so that for face-on binaries $v=\pm 1$, while for edge-on binaries $v=0$ .  
The two GW polarizations are indicated by a capital latin index, $A,B=(+,\times)$, and in the following we use the Einstein summation convention. 

By Fourier transforming Eqs.~\eqref{eq:waveform_at_source}, and using the stationary phase approximation \cite{Finn:1992xs} one finds
\begin{equation}
\tilde h_A (f)= 
\frac{e^{-i\phi_c}}{D} \chi_A (v) \tilde{k}(f;t_c, \mathcal{M}) , 
\label{help}
\end{equation}
where we have factored out explicitly the $v$, $D$ and $\phi_c$ dependence, and where
\begin{align}
\chi_+(v) & \equiv \dfrac{1+v^2}{2},
\label{eq:chiplus} &
\chi_\times(v) &\equiv -iv,
\end{align}
and
\begin{subequations}
\begin{align}
\tilde{k}(f;t_c,\mathcal{M}) &= \sqrt{\frac{5}{24\pi^{4/3}}} \dfrac{(G\mathcal M)^{5/6}}{c^{3/2}} f^{-7/6}e^{i\Psi(f)},\label{eq:ktilde} \\
\Psi(f;t_c,\mathcal{M}) &= 2\pi f t_c -\pi/4+\dfrac{3}{4}(8\pi \mathcal{M} f)^{5/3}.
\end{align}
\end{subequations}

In the detector reference frame --- denoted by a lowercase latin index --- the signal is characterized by the dimensionless strain $\tilde{h}_a$. We denote by $n_d$ is the total number of detectors, so $a=1,\ldots n_d$. The fractional change of length in an interferometer is obtained by projecting onto the detector arms with the two ``antenna patterns'' $F^+_a$ and $F^{\times}_a$ for the two polarizations \cite{1998PhRvD..58f3001J}. These depend on the sky position of the GW source (right ascension $\alpha$ and declination $\delta$), on the polarization angle $\psi$, and on time $t$. For CBC signals, $F^A_a$ are essentially constant during the entire duration of the signal, so that we can drop the $t$ dependence.  The terms in $\psi$ in the antenna response functions can be factorized using a rotation matrix \cite{cutler1994gravitational} $$R=  \left(\begin{array}{cc} 
\cos 2 \psi & \sin 2 \psi\\ 
-\sin 2 \psi & \cos 2 \psi
\end{array}\right),$$ such that,
\begin{equation}
\label{eq:antenna}
F^A_a (\alpha,\delta,\psi) = R^A_B (2 \psi) \hat{F}^B_a(\alpha,\delta),
\end{equation}
where $\hat{F}^A_a \equiv  F^A_a(\alpha,\delta,\psi=0)$. In the frequency domain, the signal in the detector reference frame can by obtained from Eqs.~\eqref{help} and \eqref{eq:antenna} as
\begin{equation}
\label{eq:facto_waveform}
\tilde{h}_a(f)= \frac{e^{-i(\phi_c+2\pi f \tau_a)}}{D} R^A_B (2 \psi) \hat{F}^B_a(\alpha,\delta)\chi_A (v) \tilde{k}(f;t_c, \mathcal{M}),
\end{equation}
where we have added a delay term $\tau_a$ relative to the time travel of the GW from the center of the geocentric frame to the detector.

From Eq.~\eqref{eq:facto_waveform}, the signal phase depends on the two {\it intrinsic parameters}\footnote{At higher PN order, the spins should be also taken into account \cite{Sathyaprakash2009}.} $\mathcal{M}$ and $t_c$.
There are also {\it extrinsic parameters}, namely the source sky coordinates $\alpha, \delta$ (that are assumed to be known) and four other parameters $D, \phi_c, v$ and $\psi$ which determine the amplitude scale of the GW signal.
These latter parameters are recovered from the measurement of the GW polarization amplitudes $+$ and $\times$. In principle, a noise-free signal observed in at least three detectors leads to a perfect measurement of the two polarizations $+$ and $\times$, and thus enables us to completely solve for these four variables. However, this is not true in practise, as the presence of noise and correlations between those parameters greatly limits the accuracy of their estimates, and the detector network may not be sensitive to both polarizations.

\subsection{Data analysis framework}

In order to understand the main sources of error on the distance $D$, we briefly summarize the canonical data analysis-framework for matched filtering.  In the following, the output (data, noise, etc) associated with the $n_d$-detector network will be denoted by $\vec{x}=(x_1,\ldots,x_{n_d})$.

The data $\vec{s}$ is modelled as a superposition of noise $\vec{\eta}$ and signal $\vec{h}(\boldsymbol{\theta})$, where $\boldsymbol{\theta}$ are the set of eight parameters introduced previously,
\begin{equation}
    \vec{s} = \vec{\eta} + \vec{h}(\boldsymbol{\theta}).
    \label{eq:super}
\end{equation}
We assume that the noise $\vec{\eta}$ is stationary, Gaussian distributed with zero mean and it is uncorrelated between different detectors\footnote{We consider detectors separated by large distances, $\gtrsim$ 1000 km so that the noise in the observable frequency bandwidth can be reasonably assumed uncorrelated.}. The likelihood for a signal $\vec{h}$ to be present in the detector data $\vec{s}$ is \cite{Jaranowski2012}
\begin{equation}
\label{eq:likelihood}
    \mathcal{L}(\vec{s}|\vec{h} (\boldsymbol{\theta}))\propto e^{-\frac{1}{2} \braket{\vec{s}-\vec{h}(\boldsymbol{\theta})|\vec{s}-\vec{h}(\boldsymbol{\theta})}},
\end{equation}
where the scalar product is defined by
\begin{equation}
\label{eq:scal_p}
    \braket{\vec{a}|\vec{b}}\equiv 
    %\sum_a \braket{a_a|b_a}= 
    4 \Re \bigg[ \int_{0}^{\infty} \dfrac{\tilde{a}^*_a(f)b_a(f)}{S_{n,a}(f)} \mathrm{d}f\bigg]
\end{equation}
where a sum over the detectors $a$ is understood. Here, `*' denotes complex conjugation, and $S_{n,a}(f)$ is the one-sided power spectral density of the detector $a$. The posterior distribution for having a signal $\vec{h}$ is given by
\begin{equation}
    P(\vec{h} (\boldsymbol{\theta})|\vec{s})\propto \mathcal{L}(\vec{s}|\vec{h} (\boldsymbol{\theta})) \: \pi(\vec{h}(\boldsymbol{\theta})),
    \label{eq:bayes}
\end{equation}
where $\pi$ represents the prior belief for the signal parameters. The uncertainties on the recovery of the parameters $\boldsymbol{\theta}$ is encoded in the posterior distributions. We can hence estimate the uncertainties on the distance evaluation by marginalizing over all the GW parameters with the exception of the distance. 

\section{Calculating the likelihood for the extrinsic parameters \label{sec:3}}

In the limit of high SNR, the Fisher matrix is commonly used to estimate the uncertainties in parameter space. However, the Fisher matrix approach has the tendency to underestimate uncertainties for correlated or degenerate parameters \cite{vallisneri2008use} as is the case for the luminosity distance and the other extrinsic parameters. Higher-order error estimate (aka ``beyond-Gaussian'') can overcome this limitation.

In the following we revisit beyond Gaussian calculations of \cite{cutler1994gravitational} to get a reasonably accurate approximation of Eq.~\eqref{eq:bayes}.

\subsection{Frame definition}

We focus on the extrinsic parameters $\boldsymbol{\beta}=(D, \phi_c, v, \psi)$, assuming a given sky position and intrinsic parameters for the GW event.
Following \cite{cutler1994gravitational} we define the amplitude factors
\begin{equation}
\label{eq:defa}
    \mathcal{A}_B=\frac{e^{-i\phi_c}}{D} R^A_B (2 \psi) \chi_A (v),
\end{equation}
so that from Eq.~\eqref{eq:facto_waveform},
\begin{equation}
\tilde{h}_a(f)= e^{- 2\pi i f \tau_a} \mathcal{A}_B \hat{F}^B_a(\vec{n}) \tilde{k}(f;t_c, \mathcal{M}),
\end{equation}
where $\vec{n}$ denotes the vector pointing to the source and is a short-hand notation for the source sky position $(\alpha, \delta)$. The scalar product $\braket{\vec{h}|\vec{h'}}$ in Eq.~\eqref{eq:scal_p} between the GW signal $h$ and the template $h'$ then becomes
\begin{equation}
\label{eq:scal_p_sec3}
\braket{\vec{h}|\vec{h'}} =\Re\left[\mathcal{A}^*_A\mathcal{A}_B'\Theta^{AB}\right]\braket{\tilde{k}|\tilde{k}}. 
\end{equation} 

We use the prime to denote template related extrinsic parameters  $\boldsymbol{\beta'}$ (template and signal have the same intrinsic parameters). From Eqs.~\eqref{eq:scal_p} and \eqref{eq:defa}, the matrix $\Theta^{AB}$ reads
\begin{equation}
	\Theta^{AB}(\vec{n})= \hat{F}_a^A(\vec{n}) \hat{F}_b^B(\vec{n}) \kappa^{ab},
\end{equation}
where, for uncorrelated detectors, the 
$n_d\times n_d$ matrix $\kappa^{ab}$ is given by
\begin{equation}
    	\kappa^{ab}=\delta^{ab} \dfrac{\int_{0}^{\infty}f^{-7/3}S_{n,a}^{-1} \mathrm{d}f}{\int_0^{\infty}f^{-7/3}S_{n,aver}^{-1}(f)\mathrm{d}f}\,.
\end{equation}
Here $S_{n,aver}$ is the average, over all the detectors, of the spectral noise densities, viz.
\begin{equation}
S_{n,aver}^{-1} (f)=n_d^{-1} \sum_{a=1}^{n_d} S_{n,a}^{-1}(f)\,
\end{equation}
so that the matrix $\kappa^{ab}$ is real and diagonal, with weights given by the noise power spectral density for each detector. Finally $\braket{\tilde{k}|\tilde{k}}$ is given by
\begin{equation}
\braket{\tilde{k}|\tilde{k}}=    4  \int_{0}^{\infty} \dfrac{|\tilde{k}(f)|^2}{S_{n,aver}(f)} \mathrm{d}f.
\end{equation}

\begin{figure*}[htp!]
    \centering
    \includegraphics[scale=0.22]{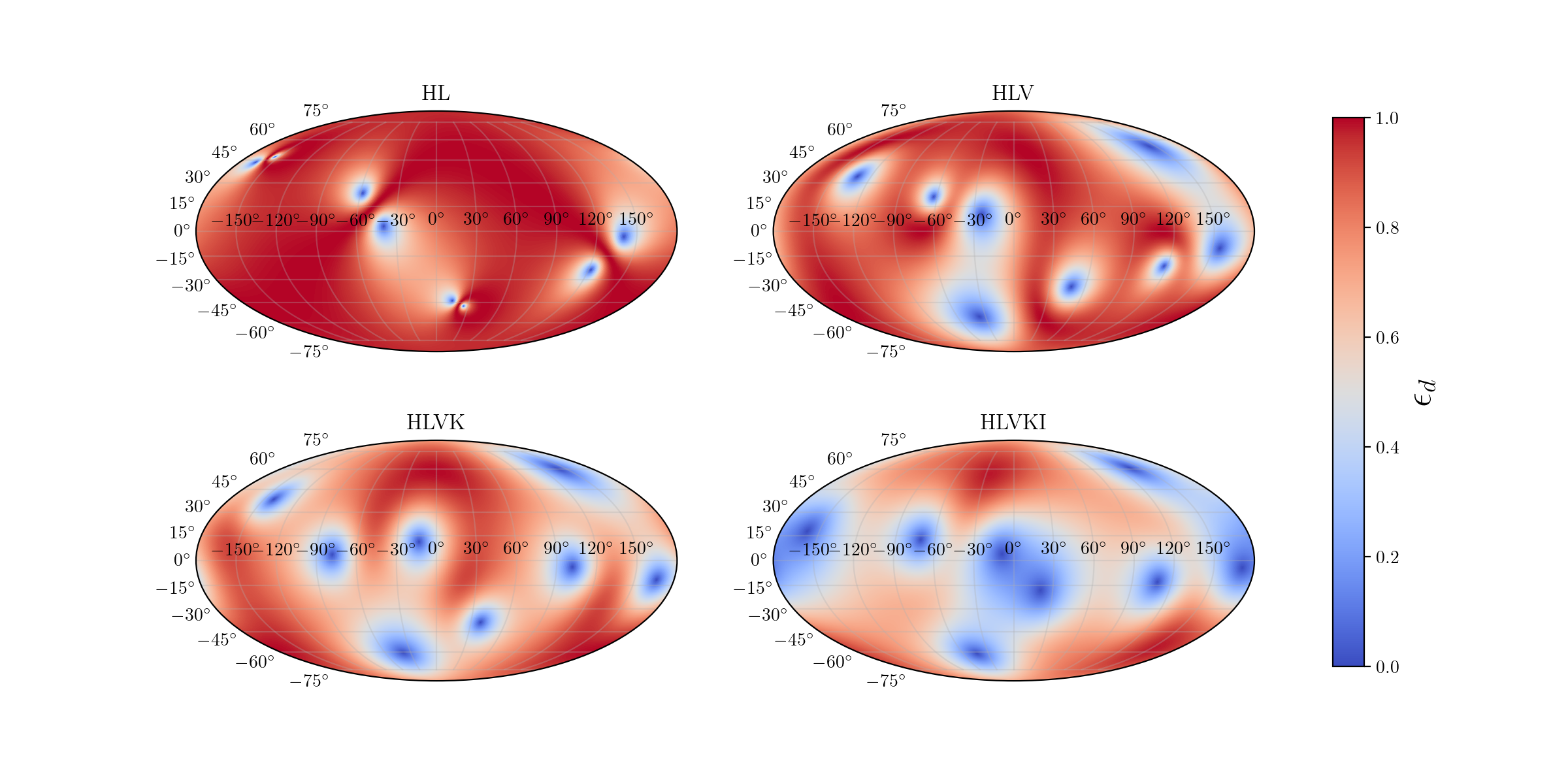}
    \caption{Value of the $\epsilon_d$ parameter with respect to the sky location for different networks of detectors assumed to be at design sensitivity. The labels ``H'', ``L'', ``V'', ``K'', ``I'' refers to the detectors present in the network, and corresponds to LIGO Hanford, LIGO Livingston, Virgo, Kagra and LIGO India, respectively. This skymap is given in equatorial coordinates at a fiducial epoch arbitrarily fixed to $t_{\rm GPS}=1187008882.43$ (Aug 17 2017 12:41:04.43 UTC)}
    \label{fig:eps_sky}
\end{figure*}

\begin{figure*}[htp!]
    \centering
    \includegraphics[scale=0.30]{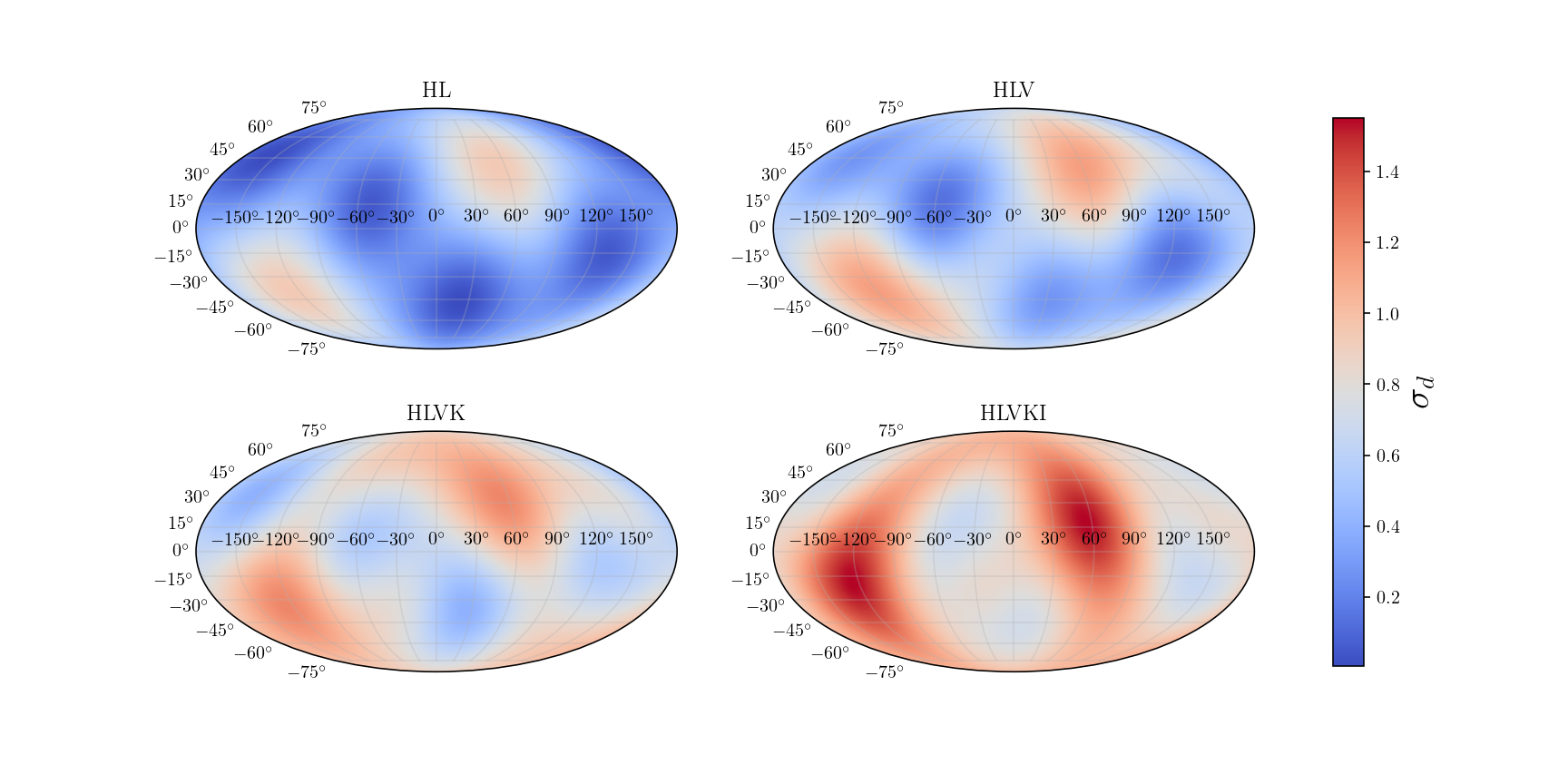}
    \caption{Value of the $\sigma_d$ parameter with respect to the sky location for different networks of detectors assumed to be at design sensitivity. The labels ``H'', ``L'', ``V'', ``K'', ``I'' refers to the detectors present in the network, and corresponds to LIGO Hanford, LIGO Livingston, Virgo, Kagra and LIGO India, respectively. This skymap is given in equatorial coordinates at a fiducial epoch arbitrarily fixed to $t_{\rm GPS}=1187008882.43$ (Aug 17 2017 12:41:04.43 UTC)}
    \label{fig:sigma_sky}
\end{figure*}

The matrix $\Theta^{AB}$ is a real non-diagonal symmetric matrix which incorporates information on position of the source in the sky, as well as geometrical information of the detector network, weighted by the detectors sensitivity $\kappa^{ab}$. This matrix is independent of the polarization angle $\psi$, and characterizes the response of the detector network to the $+$ and $\times$ polarizations with $\psi=0$.

The matrix $\Theta^{AB}$ can be diagonalized by
\begin{equation}
	\bar\Theta \equiv R(2\Delta \psi) \Theta R(-2\Delta \psi),
\end{equation}
with
\begin{equation}
   \Delta \psi=\frac{1}{4} \arctan\left[ \frac{2\Theta^{+ \times}}{\Theta^{++}-\Theta^{\times \times}}\right]\, .
   \label{eq:fedup}
\end{equation}

This is equivalent to defining effective polarizations $\bar{+}$ and $\bar{\times}$ obtained from the original $+$ and $\times$ by rotating the definition of $\psi$ by the angle $\Delta \psi$. The detector response $\bar{\Theta}^{AB}$ to $\bar{+}$ and $\bar{\times}$ takes the simple form
\begin{equation}
	\bar{\Theta}^{AB} =  \sigma_d \left(\begin{array}{cc} 
	1+\epsilon_d & 0\\ 
	0 & 1-\epsilon_d
	\end{array}\right), 
	\label{eq:thetadiag}
\end{equation} 
where $\sigma_d \equiv (\Theta^{++} + \Theta^{\times \times})/2$ and $0\leq \epsilon_d \leq 1$. Both those quantities can be also expressed in terms of the antenna patterns $\hat{F}^A_a(\vec{n})$.

The parameter $\epsilon_d$ indicates how well the effective polarization amplitudes can be measured from the detector network data \cite{cutler1994gravitational}. When $\epsilon_d=0$, $\bar{+}$ and $\bar{\times}$ can be estimated equally well. For $\epsilon_d=1$, only $\bar{+}$ can be estimated. We elaborate further on this point in Sec.~\ref{sec:discuss_degeneracy}.

Fig.~\ref{fig:eps_sky} shows the sky distribution of the parameter $\epsilon_d$ for different networks of GW detectors (we also show, for comparison, a similar plot for $\sigma_d$ in Fig.~\ref{fig:sigma_sky}). Clearly, the areas where both polarizations can be better disentangled ($\epsilon_d$ small) expand when the number of detectors increases.

\begin{figure}[h!]
    \centering
    \includegraphics[scale=1]{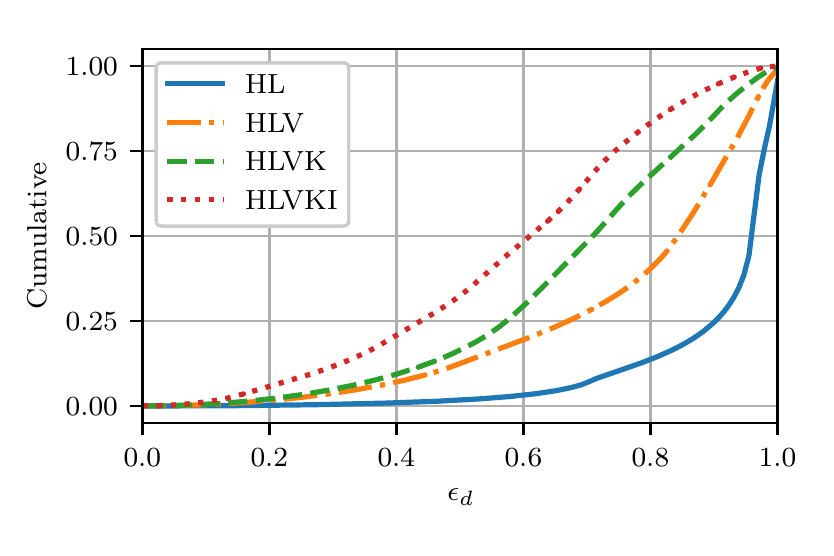}
    \caption{Cumulative density function of the number of sky points associated to an $\epsilon_{d}$ value for several detector networks at design sensitivity.}
    \label{fig:eps_den}
\end{figure}

Assuming GW sources are equally distributed in the sky, Fig.~\ref{fig:eps_den} shows the probability that the source position has an $\epsilon_d$ value lower than a certain threshold. About 80 \% of the sources have $\epsilon_d \lesssim 0.8$ for the five detector network HLVKI\footnote{The labels ``H'', ``L'', ``V'', ``K'', ``I'' refers to the detectors present in the network, and corresponds to LIGO Hanford, LIGO Livingston, Virgo, Kagra and LIGO India, respectively.} whereas this is less than 10 \% of the sources for the two detector network HL.
Using the effective polarization amplitudes $\bar{\mathcal{A}}_B = {\cal A}_B(\bar{\psi}=\psi + \Delta \psi)$, Eq.~\eqref{eq:scal_p_sec3} takes the form
\begin{equation}
\label{eq:scal_p_sec3bis}
\braket{\vec{h}|\vec{h'}} =\Re\left[\bar{\mathcal{A}}^*_A \bar{\mathcal{A}}'_B \bar{\Theta}^{AB} \right]\braket{\tilde{k}|\tilde{k}}\:. %\Re\left[\mathcal{A}^*_A\mathcal{A}_B'\Theta^{AB}\right](k|k),\.
\end{equation} 

On substituting Eq.~\eqref{eq:thetadiag} into Eq.~\eqref{eq:scal_p_sec3bis}, we find
\begin{equation}
     \braket{\vec{h}|\vec{h'}} = \frac{\sigma_d \braket{\tilde{k}|\tilde{k}}}{2 D D'} \left[ \braket{\vec{h}|\vec{h'}}_0 + \epsilon_d \braket{\vec{h}|\vec{h'}}_{\epsilon_d} \right] \label{eq:matched_new}
\end{equation}
with
\begin{subequations}
\label{eq:definitions}
\begin{align}
 \braket{\vec{h}|\vec{h'}}_0 = &(\chi_+-v)(\chi'_+-v') \cos(\varphi+2\psi_-) + \\
& (\chi_+ + v)(\chi'_+ + v') \cos(\varphi-2\psi_-) \nonumber\\
 \braket{\vec{h}|\vec{h'}}_{\epsilon_d} = & (\chi_+-v)(\chi'_++v') \cos(\varphi+2\psi_+) + \\
& (\chi_+ + v)(\chi'_+ - v') \cos(\varphi-2\psi_+) \nonumber
\end{align}
\end{subequations}
where $\chi_+$ is given in \eqref{eq:chiplus}, and 
\begin{align}
  \varphi&\equiv \phi_c-\phi_c'  &
  \psi_{\pm}&\equiv \bar{\psi} \pm \bar{\psi'}.
\end{align}

Finally, using \eqref{eq:matched_new}, the network SNR is given by
\begin{equation}
    \rho^2  \equiv \braket{\vec{h}|\vec{h}}= \rho_0^2  {\sigma_d} \Big[  (\chi_+^2 + v^2)+  \epsilon_d  (\chi_+^2 - v^2)\cos(4\bar{\psi})\Big]
    \label{eq:SNR}
\end{equation}
where, on using \eqref{eq:ktilde},
\begin{equation}
    \rho_0^2 \equiv \frac{\braket{\tilde{k}|\tilde{k}}}{D^2} = 
    \frac{1}{D^{2}}\left[ \frac{5}{6\pi^{4/3}}\frac{G{\cal M}^{5/3}}{c^3}\int_0^\infty \mathrm{d}f \dfrac{f^{-7/3}}{S_{n,aver}(f)} \right]\:.
\end{equation}
For a source that is overhead, $\sigma_d = n_d/2$. If the source is also face-on (i.e., $\chi_+ = 1$), it follows from Eq.~\eqref{eq:SNR} that $\rho^2 = n_d \rho_0^2$. Thus $\rho_0$ can be interpreted as the single-detector SNR for a face-on source located overhead \cite{cutler1994gravitational}.

\subsection{Posterior probability density}

The posterior probability density $P({\boldsymbol{\beta'}}|\vec{s})$ for the extrinsic parameters ${\boldsymbol{\beta'}}=(D', \bar{\psi'}, v', \phi'_c)$ given detector data $\vec{s}$, can be computed from the above identities. 
The source sky location and intrinsic parameters $\boldsymbol{\lambda'}=(\alpha',\delta',\mathcal{M'},t'_c)$ are known and can be factored out,
so that from Eq.~\eqref{eq:bayes} we have\footnote{When the parameters $\mathcal{M'}$ and $t'_c$ are unknown, this still remains true in the large SNR limit \cite{PhysRevD.48.4738,PhysRevD.49.1723,cutler1994gravitational}}
\beq
    P(\boldsymbol {\beta '}|\vec{s},\boldsymbol{\lambda})\propto  \mathcal{L}(\vec{s}|\boldsymbol{\beta '},\boldsymbol{\lambda}) \pi(\boldsymbol{\beta'}),
\eeq
where $\pi(\boldsymbol{\beta'})$ represents the prior probability on $\boldsymbol{\beta'}$ and is assumed to be independent of $\boldsymbol{\lambda'}$.

Using \eqref{eq:super}, the likelihood can be factorized as
\begin{equation}
    \mathcal{L}(\vec{s}|\boldsymbol{\beta'},\boldsymbol{\lambda})\propto {\rm exp}\left[-\frac{1}{2}\braket{\eta|\eta} -\frac{1}{2}\braket{h-h'|h-h'}\right].
    \label{eq:CENS}
\end{equation}
In the exponential we have neglected the terms $\braket{\eta|h-h'}$ as they are
much smaller than the terms in $\braket{h-h'|h-h'}$ in the high SNR limit.
For simplicity we drop the term $\braket{\eta|\eta}$ as it is independent of the intrinsic parameters and acts as a normalization constant. Substituting the scalar products in Eqs.~\eqref{eq:definitions} we obtain 
\begin{widetext}
\begin{multline}
\label{eq:longlong}
    \ln P(\boldsymbol{\beta'}|\vec{s})
    \propto  -\frac{\rho^2_0 \sigma_d}{2} \Big\{ (\chi_+^2 + v^2) + d^{-2}(\chi_+'^2 + v'^2) - d^{-1}(\chi_+ - v)(\chi'_+ - v') \cos(\varphi + 2\psi_-) - d^{-1}(\chi_+ + v)(\chi'_+ + v') \cos(\varphi - 2\psi_-) \\
+ \epsilon_d \big[ (\chi_+^2 - v^2) \cos(4\bar{\psi}) + d^{-2}(\chi_+'^2 - v'^2)\cos(4\bar{\psi'})  - d^{-1}(\chi_+ - v)(\chi'_+ + v') \cos(\varphi + 2\psi_+) - d^{-1}(\chi_+ + v)(\chi'_+ - v') \cos(\varphi - 2\psi_+) \big] \Big\} \\
   + \ln \pi(\boldsymbol{\beta'}),
\end{multline}
\end{widetext}
where $d \equiv D'/D$ is the ratio of the template distance to the true distance.

In order to determine the luminosity distance uncertainty, we need to marginalize Eq.~\eqref{eq:longlong} over the extrinsic parameters $v',\psi'$ and $\phi_c'$. Unfortunately, this cannot be done analytically as this expression depends on the true extrinsic parameters which are not known. For this reason we present several approximations for Eq.~\eqref{eq:longlong} in the following section.

\section{Marginalized distance posterior \label{sec:4}}

We first begin by briefly discussing the effect of $\epsilon_d$ on the likelihood probability density function.

\subsection{Relation of $\epsilon_d$ to the degeneracy of the estimation problem}
\label{sec:discuss_degeneracy}

Using the definitions of the dominant polarization frame in Eq.~\eqref{eq:scal_p_sec3bis} as well as Eq.~\eqref{eq:CENS}, the likelihood can be written in the more compact form
\begin{equation}
     \ln \mathcal{L}\propto \sigma_d \rho_0^2(1+\epsilon_d) |\bar{\mathcal{A}}_+ - \bar{\mathcal{A}}'_+|^2 + \sigma_d \rho_0^2(1-\epsilon_d) |\bar{\mathcal{A}}_\times - \bar{\mathcal{A}}'_\times|^2,
    \label{eq:liklik}
\end{equation}
 where in this expression (and with a slight abuse of notation) the amplitudes should be understood to be expressed in terms of the normalised distance $d$.
In the $\bar{\mathcal{A}}'_{+/\times}$ plane, the likelihood is a bivariate Gaussian distribution centered on the true values $\bar{\mathcal{A}}_{+/\times}$. This is due to the fact that we are neglecting noise terms in the likelihood scalar products in Eq.~\ref{eq:CENS}.
The $\epsilon_d$ parameter determines the variances for this Gaussian function along the directions $\bar{+}$ and $\bar{\times}$, namely $\bar{\sigma}^2_{+}=1/[2\sigma_d \rho_0^2(1+\epsilon_d)]$ and $\bar{\sigma}^2_{\times}=1/[2 \sigma_d\rho_0^2(1-\epsilon_d)]$.

It follows that the level curves of constant likelihood are ellipses with semi-minor and semi-major axes proportional to $\bar{\sigma}^2_{+}$ and $\bar{\sigma}^2_{\times}$.
The relative length of the curve identifying the constant likelihood surfaces is proportional to the parameter space volume that would give such likelihood values. The curve length is $\propto 1/\rho_0\sqrt{\sigma_d(1-\epsilon_d^2)}$ and thus increases with $\epsilon_d$ but decreases with the SNR $\rho_0$.

We show in Appendix \ref{AppA} that, when $\epsilon_d < 1$, ${\cal L}$ has a unique global maximum, i.e., $\bar{\mathcal{A}}'_{A} = \bar{\mathcal{A}}_{A}$ which corresponds to the template with $d=1$, $v'=v$ and $\bar{\psi}'=\bar{\psi}$. When $\epsilon_d \rightarrow 1$, $\sigma_\times \rightarrow \infty$ meaning that the likelihood becomes flat along the $\bar{\cal{A}}_\times$ axis --- the detector network is blind to the $\bar{\times}$ polarization. As a result the maximum likelihood is not a single point but the line $\bar{\mathcal{A}}_{+} = \bar{\mathcal{A}}'_{+}$. Many solutions for $d,v',\bar{\psi}'$ or equivalently many physical templates (given in Appendix \ref{AppA}) maximise the likelihood.

Phrased differently, the estimation amounts to resolving the system of two complex-valued equations $\bar{\mathcal{A}}'_{A} = \bar{\mathcal{A}}_{A} + \delta \bar{\mathcal{A}}_{A}$ for $A = \{+, \times\}$ and where $\delta \bar{\mathcal{A}}_{A}$ is a random perturbation of the same order of the standard deviation $\bar{\sigma}_{A}$. 
The system corresponds to four real-valued equations which allows one to solve for the four real-valued unknowns $(D', \phi'_c, v', \psi')$. In the large SNR limit $\rho_0 \rightarrow \infty$ and $\epsilon_d < 1$, the perturbation terms vanish and the system can be exactly resolved. When $\epsilon_d = 1$, the perturbation in the second equation is infinite. This equation cannot be solved and the system is under-determined, thus leading to many degenerate solutions. The degeneracy remains when $\epsilon_d$ is  close to 1, i.e., when the amplitude of the perturbation is comparable to the $\bar{\mathcal{A}}_{\times}$ polarization amplitude. This happens when
\begin{equation}
    \epsilon_d \gtrsim 1- \frac{1}{\rho^2_0}.
\end{equation}

\subsection{Approximations of the distance posterior}

We now present different expressions for the {\it posterior probability density} obtained in Eq.~\eqref{eq:longlong} that can
be used to estimate the luminosity distance uncertainty,
\begin{equation}
    \left(\frac{\Delta D}{D}\right) \equiv  \frac{\sqrt{\int {\rm d} d \, P(d|\vec{s})(d-\bar{d})^2} }{\int {\rm d}d \, P(d|\vec{s}) d}
\end{equation}
where the marginalised posterior is given by
\begin{equation}
   P(d|\vec{s})=\int{\rm d}v' {\rm d}\psi' {\rm d}\phi_c' P(d,v',\psi',\phi_c'|\vec{s}).
\end{equation}

\subsubsection{Exact marginalized posterior for $\epsilon_d=0$}
\label{subsec:eps0}

We first study the posterior when $\epsilon_d =0$, i.e., when the detector network is equally sensitive to the two effective polarizations. We obtain an exact analytical expression that can be easily evaluated numerically. This calculation is of limited use as it is obtained for a special case. It is, however, an exact expression that can serve to validate the other approximations we do.

Assuming a prior $\pi(d,\psi',v',\phi_c')= d^2$ \cite{2015PhRvD..91d2003V}, the posterior probability distribution in Eq.~\eqref{eq:longlong} can be marginalised over $\psi'$ and $\phi_c'$ to obtain
%Let $\zeta_{\pm}=\varphi \pm 2\psi_-$ and . Then . We will assume 
\begin{multline}
    P(d,v'|\vec{s}) \propto d^2 + \exp[-\frac{\rho_0^2 \sigma_d}{2} (\chi_+^2 + v^2) + d^{-2}(\chi_+'^2 + v'^2)] \\
    \times \int_{0}^{2 \pi} e^{z_- \cos(\zeta_+)} d\zeta_+ \int_{0}^{2 \pi} e^{z_+ \cos(\zeta_-)} d\zeta_-
\end{multline}
with $\zeta_{\pm}=\varphi \pm 2\psi_-$ and  $z_\pm=\frac{\rho_0^2 \sigma_d}{2d}(\chi_+ \pm v)(\chi'_+ \pm v')$. This reduces to

\begin{equation}
    P(d,v'|\vec{s}) \propto K_+ (d,v'|\vec{s}) K_- (d,v'|\vec{s}),
    \label{eq:bessel_posterior}
\end{equation}
where
\begin{multline}
  K_{\pm} (d,v'|\vec{s}) = d\: I_0(z_{\pm}) \\
  \times \exp\left(-\frac{\rho_0^2 \sigma_d}{4}[(\chi_+^2 + v^2) + d^{-2}(\chi_+'^2 + v'^2)]\right)
\end{multline}
where $I_0$ denotes the modified Bessel function.  

Eq.~\eqref{eq:bessel_posterior} does not depend on $\psi$ and $\phi_c$. This is a consequence of the fact that, when $\epsilon_d=0$, the two polarizations $\bar{+}$ and $\bar{\times}$ can be differentiated regardless of the values of $\psi$ and $\phi_c$.

\subsubsection{Cutler and Flanagan approximation}

We now obtain an approximant for the more general case, $\epsilon_d < 1$.  Following \cite{cutler1994gravitational}, we expand Eq.~\eqref{eq:longlong} to the 0-th order in $\bar{\psi}'$ and $\phi'_c$ around the true values $\bar{\psi}$ and $\phi_c$. For $\epsilon_d < 1$, this corresponds to an expansion about the unique global maximum of the posterior. The result is
\begin{multline}
  P(d,v'|\vec{s}) \propto \pi (d,v') \\
  \times \exp \big(-\frac{\rho_0^2 \sigma_d}{2} [(1-\epsilon_d   \cos 4\bar{\psi} ) (v-d^{-1}v')^2 \\
    +  (1+\epsilon_d   \cos 4\bar{\psi} )(\chi_+-d^{-1} \chi_+')^2] \big).
 \label{eq:CF-again1}
\end{multline}

\begin{figure}[h!]
    \centering
    \includegraphics[scale=0.4]{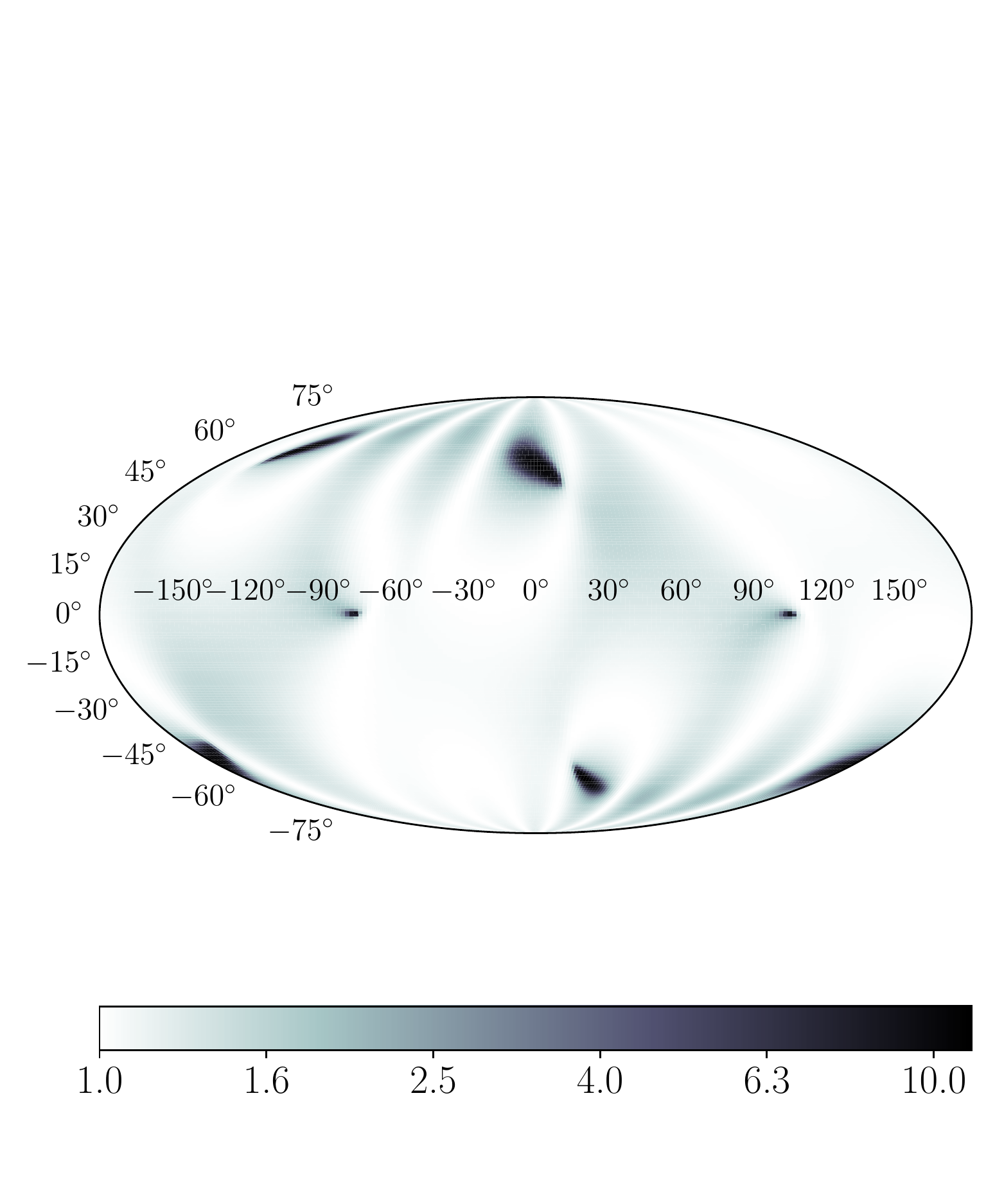}
    \caption{Ratio of the relative uncertainty on the luminosity distance $\Delta D/D$ predicted in \cite{cutler1994gravitational} and using the corrected expression in Eq.~\eqref{eq:CF-again1} for the HLV detector network. The two calculations differ by a factor up to 10 in certain parts of the sky. We assume a global SNR of 33, and use the same sky coordinates as in Fig.~\ref{fig:eps_sky}.}
    \label{fig:CF_correction}
\end{figure}

This equation corrects Eq.~(4.57) of \cite{cutler1994gravitational} which has an error. Fig.~\ref{fig:CF_correction} shows this is a significant correction reaching a factor of ten times in the predicted distance uncertainty for sky locations with $\epsilon_d \sim 1$ in the case of the HLV detector network. The approximant Eq.~\eqref{eq:CF-again1} is computationally fast to compute and can thus be evaluated easily for many sky locations and detector network. 

The final posterior $P(d,v'|\vec{s})$ in Eq.~(\ref{eq:CF-again1}) is the product of two Gaussian functions, respectively associated to the $\bar{\times}$ and $\bar{+}$ polarizations. The variance is $1/(1\pm\epsilon_d \cos(4\bar{\psi}))$, respectively. For $\epsilon_d=0$ the posterior is independent of $\bar{\psi}$, consistently to Sec.~\ref{subsec:eps0}). We verify in Sec.~\ref{sec:5} that the above approximation coincides with the exact expression obtained in Sec.~\ref{subsec:eps0} for the case $\epsilon_d=0$.

\begin{figure*}[htp!]
    \centering
    \includegraphics[scale=0.45]{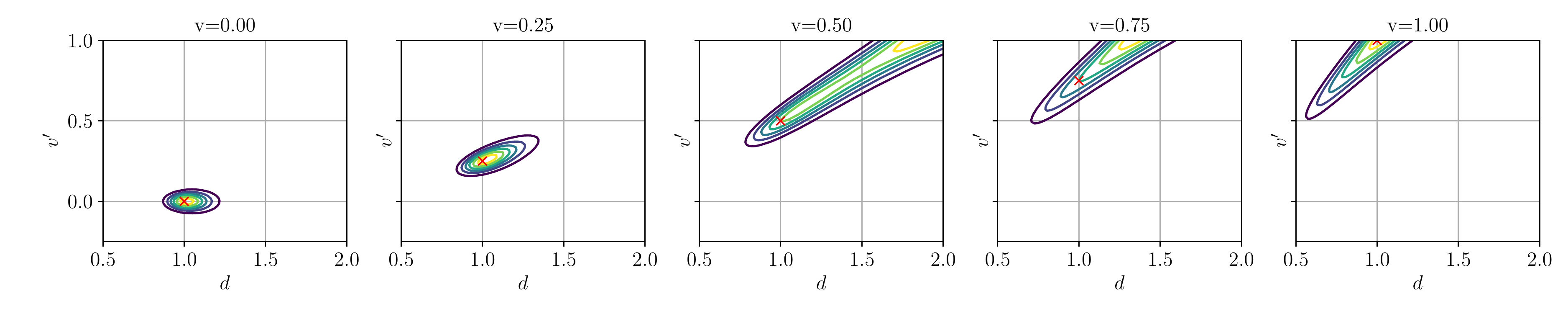}
    \caption{Posterior probability density for different values $v$ in the $d,v'$ plane predicted by the approximant in Eq.~\eqref{eq:CF-again1} for a signal with SNR=20 and $\epsilon_d=0.2$. The red cross corresponds to the maximum of the {\it likelihood} which, as discussed in Sec.~\ref{sec:3} occurs when $v'=v$ and $d=1$.}
    \label{fig:effectbessel}
\end{figure*}

Fig.~\ref{fig:effectbessel} shows the shape of the posterior in the $d,v'$ plane for $\epsilon_d=0.2$ and different values of $v$. The posterior is symmetric for $v=0$ while increasing $v$, it becomes asymmetric. This results in tails for the marginal posterior of the luminosity distance. 

\begin{figure}[htp!]
    \centering
\includegraphics[scale=1]{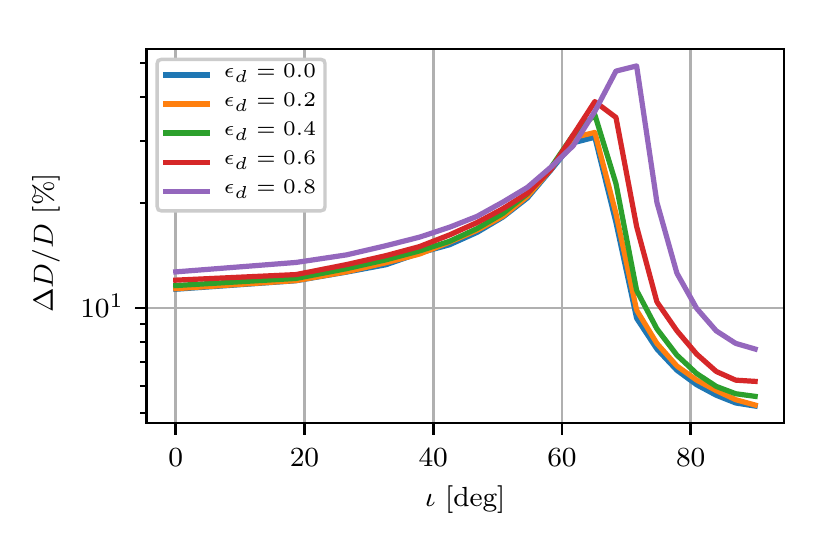}
    \caption{Relative uncertainty $\Delta D/D$ predicted by Eq.~\eqref{eq:longlong} versus the inclination angle of the binary $\iota$ for a signal with SNR=20. The different lines indicates sky locations with different $\epsilon_d$. The maximum uncertainty increases with $\epsilon_d$.}
    \label{fig:unc_wrt}
\end{figure}

These tails lead to larger errors in $D$ for large $v$ as seen in Fig.~\ref{fig:unc_wrt} where we show $\Delta D/D$ with respect to the binary inclination parameter $v$, where we have used the approximant in Eq.~\eqref{eq:longlong}, and fixed a detected SNR of 20 for a HLV network at design sensitivity,
The evolution shown in Fig.~\ref{fig:unc_wrt} for the luminosity distance is consistent to that of Fig.~1 of \cite{2018PhRvL.121b1303V} obtained by a Bayesian analysis of simulated data.
The relative distance error increases from face-on-binaries up to a maximum  for $\iota \sim 60$ deg. The maximum uncertainty increases with $\epsilon_d$ and the position of the maximum is also a function of $\epsilon_d$. Indeed, increasing $\epsilon_d$ corresponds to moving the source to sky locations in which the detector network is less capable of distinguishing the two polarizations.

\begin{figure}[htp!]
    \centering
\includegraphics[scale=1]{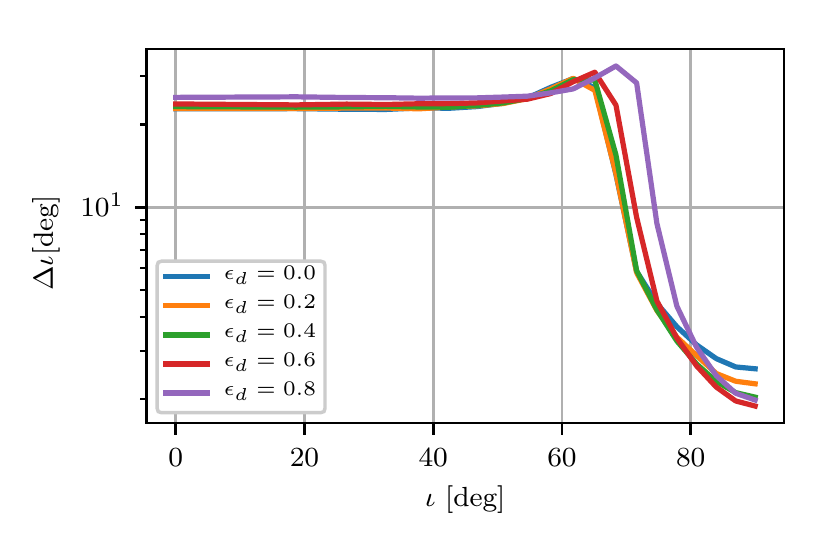}
    \caption{Uncertainty $\Delta \iota$ predicted by Eq.~\eqref{eq:longlong} versus the inclination angle of the binary $\iota$ for a signal with SNR=20. The different lines indicates sky locations with different $\epsilon_d$. The maximum uncertainty increases with $\epsilon_d$.}
    \label{fig:unc_wrt_iota}
\end{figure}

In Fig.~\ref{fig:unc_wrt_iota} we show the error budget for the determination of the inclination angle $\iota$. This appears to be consistent with Fig.~1 of \cite{2018arXiv180705226C}, predicting an error budget of about $\sim 20$ deg if the luminosity distance of the source is not constrained by any independent measure (e.g.~electromagnetic observations). 

\begin{figure*}[htp!]
    \centering
    \includegraphics[scale=0.5]{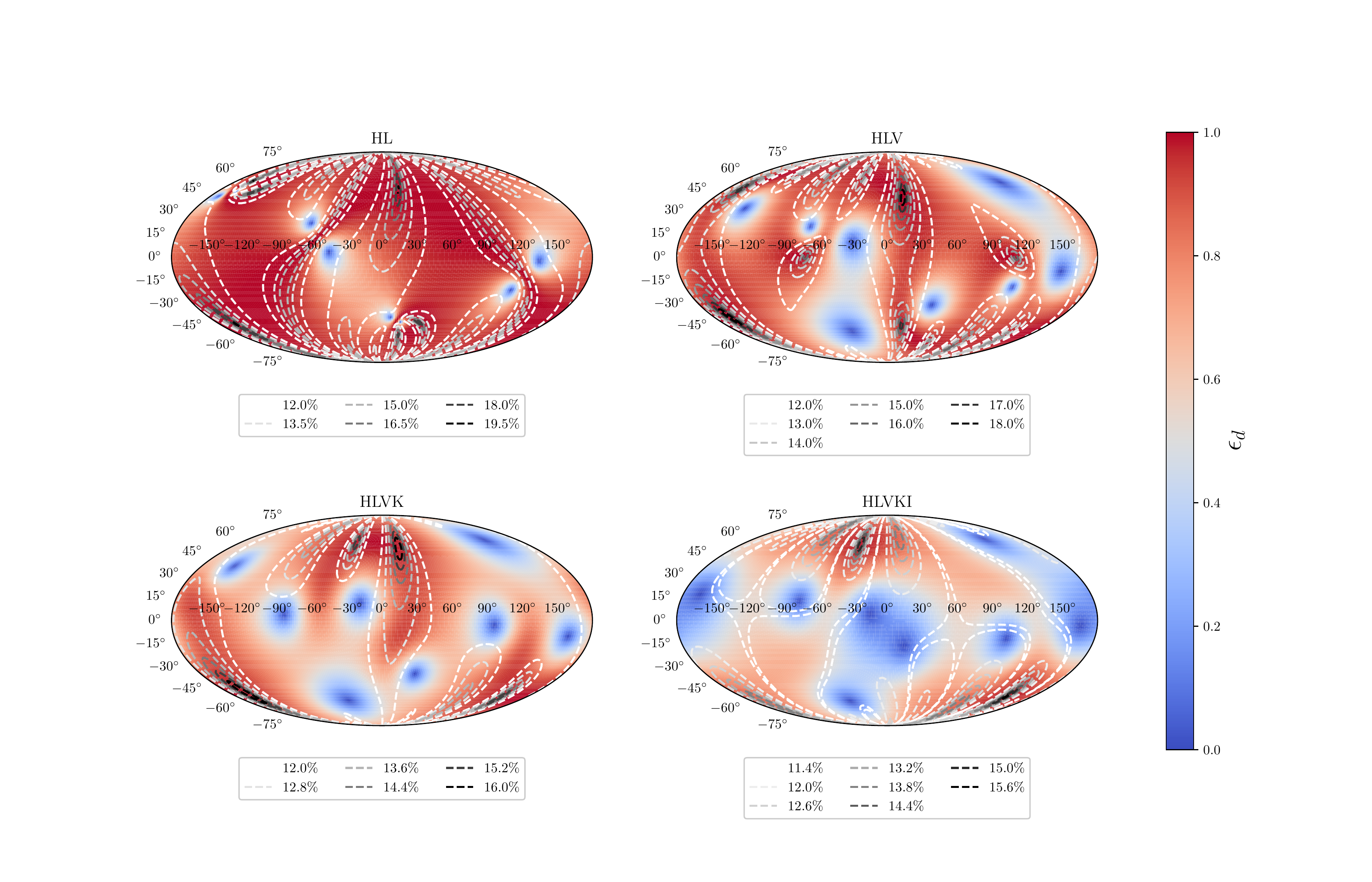}
    \caption{Level curves (gray-scaled dashed lines) for the luminosity distance accuracy measured as the posterior standard deviation from Eq.~\eqref{eq:CF-again1} over the true luminosity distance. We assume a global SNR of 33, and use the same sky coordinates as in Fig.~\ref{fig:eps_sky}. The inclination is fixed to the most probable inclination for detectable binaries, namely $v=0.85$ \cite{Schutz_2011}. There is a clear correlation with the value of $\epsilon_d$, shown with a colored mesh plot.}
    \label{fig:big_sim}
\end{figure*}

Fig.~\ref{fig:big_sim} superimposes the expected luminosity distance error with $\epsilon_d$ for several detector networks at design sensitivity. The inclination is fixed to the most probable inclination for detectable binaries, namely $v=0.85$ \cite{Schutz_2011}. 
As expected, sky patches with low $\epsilon_d$ values have more accurate measurements of the luminosity distance. 
However, has shown in Figs.~\ref{fig:unc_wrt},\ref{fig:unc_wrt_iota}, \ref{fig:big_sim}, the relative error for the luminosity distance and $\cos \iota$ has a weak scaling with $\epsilon_d$, hence even sky locations with smaller $\epsilon_d$ can not reduce the $d$-$\cos \iota$ correlations.
These result is consistent with \cite{2019ApJ...877...82U} in which a similar framework have been studied without taking into account the different sensitivities of the detectors.
On the other hand, five detector network allows us to achieve only a $10\%$ accuracy over the majority of the sky. 
Crucially, for $\epsilon_d \approx 1$, this approximant is no longer reliable since the likelihood is strongly degenerate as shown in App.~\ref{AppA} and discussed in Sec.~\ref{sec:5}.

\section{Simulations and validation\label{sec:5}}

In order to validate our predictions, we have carried out simulations of 63 binary neutron star mergers in simulated Advanced LIGO (Hanford and Livingston) \cite{2015CQGra..32g4001L} and Virgo \cite{2012arXiv1202.4031M} data at design sensitivities, using several sky locations with different values of $\epsilon_d$. We compute the posterior probability on the luminosity distance using the LALInference toolkit \cite{2015PhRvD..91d2003V, lalsuite}, by fixing the sky position for the GW events. We have used \texttt{IMRPhenomPv2} model \cite{PhysRevLett.113.151101,PhysRevD.91.024043}. The intrinsic parameters of the template, i.e., the chirp mass, mass ratio and merger time have been left to vary together with the extrinsic parameters $d,v',\psi',\phi_c'$.
The $63=9 \times 7$ injections with SNR $\rho=33$ are divided into 9 {\it sets}, labelled from ``A'' to ``H'', and 7 {\it series}. Each set has a fixed sky position (right ascension and declination) and detection epoch: see Table \ref{tab:seriesparameter} for more details.

\begin{table}[h!]
\begin{tabular}{l|cccccc}
Set& $\alpha$ [deg]& $\delta$ [deg]& $\sigma_d$& $\epsilon_d$ & {\cal C} & $\Delta \psi$[rad] \\
\hline
A&227.75&33.96&0.34&0.03& 1.05&0.27 \\
B&322.89&-55.04&0.68&0.01& 1.01& {0.02} \\
C&141.62&55.05&0.68&0.01& 1.03&-0.81 \\
D&48.66&-30.00&0.37&0.21& 1.54&-0.43 \\
E&287.57&37.93&0.30&0.79& 8.56&-0.82 \\
F&193.88&-50.00&0.87&1.00& 7283&-0.66 \\
G&288.65&0.09&0.30&1.00& 6201&0.09 \\
H&108.83&0.09&0.30&1.00& 8394&-0.09 \\
I&13.70&50.00&0.87&1.00& 8626&-0.92 \\
\end{tabular}
\caption{\label{tab:seriesparameter} The table contains the sky locations used for the simulated injections. First and second columns: right ascension and declination of the injection (epoch $t_{\rm GPS}=1187008582.0$, Aug 17 2017 12:36:04 UTC). Third and fourth columns: the corresponding values of $\sigma_d$ and $\epsilon_d$, see Eq.~\eqref{eq:thetadiag}. Fifth column: conditioning number $C$ of the matrix $\Theta$, defined as the ratio of its eigenvalues, namely $C \equiv (1+\epsilon_d)/(1-\epsilon_d)$. Sixth column: rotation angle $\Delta \psi$ in Eq.~\eqref{eq:fedup}. The injections are performed with a fixed SNR=33}
\end{table}

Each injection set is divided in 7 different series labelled from 0 to 6. Each series has a different combination of the extrinsic parameters $v,\psi$. For the series, the combinations of the chosen extrinsic parameters are given in Table~\ref{tab:series_ext}.

\begin{table}[h!]
\begin{tabular}{l|cc}
Series& $v$ & $\psi$ [deg]\\
\hline
0&1.00&0 \\
1&0.87&0\\
2&0.50&0 \\
3&0&0 \\
4&0&29.79 \\
5&0&60.16 \\
6&0&90.00\\
\end{tabular}
\caption{\label{tab:series_ext} The table contains the extrinsic parameters for each injection series. First column: cosine of the inclination of the orbital plane with respect to the line of sight. Second column: Polarization angle $\psi$ in degrees.}
\end{table}

The results of our simulations are summarised in Table \ref{eq:full_table}, where we report the mean $\bar{d}$ and the variance $\sigma^2$, together the skewness $\gamma$ obtained for the posterior distribution of the normalized luminosity distance $d$. The marginal posterior distribution for the normalized distance $d$ obtained with LALInference is compared with the predictions obtained with the CF approximation, as well as through a direct numerical integration, by sampling from the full posterior in Eq.~\eqref{eq:longlong} using MCMC \footnote{The MCMC marginalization was carried out using a Parallel tempered ensemble MCMC \cite{2013PASP..125..306F} with 250 walkers, 3 temperatures logarithmically spaced between 1 and 10 and 10000 samples, taking into account the integrated autocorrelation time.}. The MCMC marginalization of Eq.~\eqref{eq:longlong} is performed over two additional variables with respect to the CF approximation. It is thus expected to lead to larger credible intervals that better matches with those of LALInference.

LALInference is able to recover the SNR close to the injected value $\sim 33$.

\begingroup
\squeezetable
\begin{table*}
\caption{\label{eq:full_table}This table reports three statistical the mean $\bar{d}$, variance $\sigma^2$ and skewness $\gamma$ for the posterior probability density function of the normalized luminosity distance $d$ of each injection. First, second and third columns: posterior obtained with LALInference. Fourth, fifth and sixth columns: posterior obtained with the CF approximant. Seventh, eighth and ninth columns: posterior obtained with the MCMC marginalization of Eq.~\eqref{eq:longlong}.}
\begin{tabular}{l|ccc|ccc|ccc}
\textbf{Injection}& $\bar{d}_{\rm{LAL}}$ & $\sigma^2_{\rm{LAL}}$ & $\gamma_{\rm{LAL}}$ & $\bar{d}_{\rm{CF}}$ & $\sigma^2_{\rm{CF}}$ & $\gamma_{\rm{CF}}$ & $\bar{d}_{\rm{MC}}$ & $\sigma^2_{\rm{MC}}$ & $\gamma_{\rm{MC}}$ \\ 
\hline
A-0 & 0.82 & 1.66$\cdot 10^{-2}$ & -0.11& 0.89 & 8.84$\cdot 10^{-3}$ & -0.43& 0.88 & 8.29$\cdot 10^{-3}$ & -0.43 \\ 
A-1 & 0.88 & 1.98$\cdot 10^{-2}$ & 0.05& 1.01 & 1.20$\cdot 10^{-2}$ & -0.40& 1.01 & 1.06$\cdot 10^{-2}$ & -0.44 \\ 
A-2 & 1.07 & 2.00$\cdot 10^{-2}$ & 2.20& 1.05 & 1.15$\cdot 10^{-2}$ & 1.90& 1.10 & 2.47$\cdot 10^{-2}$ & 2.05 \\ 
A-3 & 0.99 & 8.99$\cdot 10^{-4}$ & 0.20& 1.01 & 9.35$\cdot 10^{-4}$ & 0.19& 1.01 & 9.38$\cdot 10^{-4}$ & 0.19 \\ 
A-4 & 1.13 & 1.46$\cdot 10^{-3}$ & 0.21& 1.01 & 1.19$\cdot 10^{-3}$ & 0.21& 1.03 & 6.36$\cdot 10^{-4}$ & 0.95 \\ 
A-5 & 1.00 & 9.10$\cdot 10^{-4}$ & 0.21& 1.01 & 9.11$\cdot 10^{-4}$ & 0.18& 1.01 & 9.13$\cdot 10^{-4}$ & 0.18 \\ 
A-6 & 1.00 & 9.07$\cdot 10^{-4}$ & 0.16& 1.01 & 9.16$\cdot 10^{-4}$ & 0.18& 1.01 & 9.21$\cdot 10^{-4}$ & 0.17 \\ 
B-0 & 0.57 & 3.84$\cdot 10^{-3}$ & 2.39& 0.89 & 8.77$\cdot 10^{-3}$ & -0.43& 0.88 & 7.90$\cdot 10^{-3}$ & -0.48 \\ 
B-1 & 0.89 & 1.65$\cdot 10^{-2}$ & -0.19& 1.01 & 1.19$\cdot 10^{-2}$ & -0.40& 1.01 & 1.02$\cdot 10^{-2}$ & -0.48 \\ 
B-2 & 1.04 & 1.29$\cdot 10^{-2}$ & 2.53& 1.06 & 1.26$\cdot 10^{-2}$ & 1.93& 1.11 & 2.77$\cdot 10^{-2}$ & 1.94 \\ 
B-3 & 0.97 & 7.99$\cdot 10^{-4}$ & 0.16& 1.01 & 8.37$\cdot 10^{-4}$ & 0.18& 1.01 & 8.05$\cdot 10^{-4}$ & 0.08 \\ 
B-4 & 1.00 & 9.19$\cdot 10^{-4}$ & 0.19& 1.01 & 9.25$\cdot 10^{-4}$ & 0.18& 1.01 & 9.27$\cdot 10^{-4}$ & 0.18 \\ 
B-5 & 1.04 & 1.07$\cdot 10^{-3}$ & 0.21& 1.01 & 9.84$\cdot 10^{-4}$ & 0.19& 1.01 & 9.59$\cdot 10^{-4}$ & 0.26 \\ 
B-6 & 1.01 & 9.41$\cdot 10^{-4}$ & 0.17& 1.01 & 9.17$\cdot 10^{-4}$ & 0.18& 1.01 & 9.21$\cdot 10^{-4}$ & 0.20 \\ 
C-0 & 0.86 & 8.83$\cdot 10^{-3}$ & -0.43& 0.89 & 8.76$\cdot 10^{-3}$ & -0.43& 0.88 & 8.23$\cdot 10^{-3}$ & -0.44 \\ 
C-1 & 1.05 & 1.30$\cdot 10^{-2}$ & -0.43& 1.01 & 1.27$\cdot 10^{-2}$ & -0.40& 1.01 & 1.13$\cdot 10^{-2}$ & -0.41 \\ 
C-2 & 1.32 & 5.49$\cdot 10^{-2}$ & 0.40& 1.05 & 1.05$\cdot 10^{-2}$ & 1.84& 1.09 & 2.25$\cdot 10^{-2}$ & 2.17 \\ 
C-3 & 1.04 & 1.07$\cdot 10^{-3}$ & 0.20& 1.01 & 1.02$\cdot 10^{-3}$ & 0.19& 1.01 & 9.94$\cdot 10^{-4}$ & 0.27 \\ 
C-4 & 0.98 & 8.27$\cdot 10^{-4}$ & 0.17& 1.01 & 8.84$\cdot 10^{-4}$ & 0.18& 1.01 & 8.90$\cdot 10^{-4}$ & 0.16 \\ 
C-5 & 1.03 & 1.02$\cdot 10^{-3}$ & 0.20& 1.01 & 9.55$\cdot 10^{-4}$ & 0.19& 1.01 & 9.57$\cdot 10^{-4}$ & 0.20 \\ 
C-6 & 1.05 & 1.12$\cdot 10^{-3}$ & 0.18& 1.01 & 9.88$\cdot 10^{-4}$ & 0.19& 1.01 & 9.75$\cdot 10^{-4}$ & 0.22 \\ 
D-0 & 0.82 & 1.24$\cdot 10^{-2}$ & -0.30& 0.89 & 8.74$\cdot 10^{-3}$ & -0.44& 0.88 & 8.27$\cdot 10^{-3}$ & -0.44 \\ 
D-1 & 1.02 & 1.41$\cdot 10^{-2}$ & -0.42& 1.01 & 1.26$\cdot 10^{-2}$ & -0.40& 1.01 & 1.14$\cdot 10^{-2}$ & -0.43 \\ 
D-2 & 0.92 & 3.39$\cdot 10^{-3}$ & 0.60& 1.06 & 1.18$\cdot 10^{-2}$ & 1.90& 1.06 & 7.47$\cdot 10^{-3}$ & 0.35 \\ 
D-3 & 1.01 & 1.02$\cdot 10^{-3}$ & 0.19& 1.01 & 9.48$\cdot 10^{-4}$ & 0.19& 1.01 & 1.00$\cdot 10^{-3}$ & 0.19 \\ 
D-4 & 1.01 & 9.55$\cdot 10^{-4}$ & 0.15& 1.01 & 8.72$\cdot 10^{-4}$ & 0.18& 1.01 & 8.80$\cdot 10^{-4}$ & 0.17 \\ 
D-5 & 0.98 & 8.18$\cdot 10^{-4}$ & 0.21& 1.01 & 8.40$\cdot 10^{-4}$ & 0.18& 1.01 & 8.51$\cdot 10^{-4}$ & 0.15 \\ 
D-6 & 1.05 & 1.17$\cdot 10^{-3}$ & 0.22& 1.01 & 9.82$\cdot 10^{-4}$ & 0.19& 1.01 & 9.58$\cdot 10^{-4}$ & 0.35 \\ 
E-0 & 0.79 & 2.36$\cdot 10^{-2}$ & 0.15& 0.87 & 1.26$\cdot 10^{-2}$ & -0.57& 0.86 & 1.12$\cdot 10^{-2}$ & -0.47 \\ 
E-1 & 0.98 & 1.73$\cdot 10^{-2}$ & -0.39& 1.00 & 1.71$\cdot 10^{-2}$ & -0.56& 0.99 & 1.47$\cdot 10^{-2}$ & -0.48 \\ 
E-2 & 1.41 & 1.12$\cdot 10^{-1}$ & 0.16& 1.27 & 8.65$\cdot 10^{-2}$ & 0.81& 1.49 & 9.72$\cdot 10^{-2}$ & -0.18 \\ 
E-3 & 0.98 & 8.50$\cdot 10^{-4}$ & 0.22& 1.01 & 8.70$\cdot 10^{-4}$ & 0.18& 1.01 & 8.97$\cdot 10^{-4}$ & 0.11 \\ 
E-4 & 1.04 & 1.67$\cdot 10^{-3}$ & 0.06& 1.01 & 9.59$\cdot 10^{-4}$ & 0.19& 1.02 & 2.31$\cdot 10^{-3}$ & 0.20 \\ 
E-5 & 1.07 & 2.48$\cdot 10^{-3}$ & 0.09& 1.01 & 9.51$\cdot 10^{-4}$ & 0.19& 1.01 & 1.79$\cdot 10^{-3}$ & 0.10 \\ 
E-6 & 0.98 & 8.76$\cdot 10^{-4}$ & 0.18& 1.01 & 8.99$\cdot 10^{-4}$ & 0.18& 1.01 & 9.33$\cdot 10^{-4}$ & 0.14 \\ 
F-0 & 0.76 & 2.35$\cdot 10^{-2}$ & -0.64& 0.86 & 1.36$\cdot 10^{-2}$ & -0.62& 0.78 & 2.73$\cdot 10^{-2}$ & -0.63 \\ 
F-1 & 0.90 & 4.05$\cdot 10^{-2}$ & -0.70& 0.99 & 1.93$\cdot 10^{-2}$ & -0.61& 0.90 & 3.77$\cdot 10^{-2}$ & -0.58 \\ 
F-2 & 1.56 & 1.14$\cdot 10^{-1}$ & -0.62& 1.41 & 1.12$\cdot 10^{-1}$ & 0.16& 1.54 & 1.10$\cdot 10^{-1}$ & -0.62 \\ 
F-3 & 6.01 & 1.60 & -0.69& 1.01 & 8.79$\cdot 10^{-4}$ & 0.18& 6.09 & 1.77 & -0.64 \\ 
F-4 & 2.19 & 2.17$\cdot 10^{-1}$ & -0.63& 1.01 & 9.34$\cdot 10^{-4}$ & 0.20& 1.62 & 1.19$\cdot 10^{-1}$ & -0.57 \\ 
F-5 & 1.65 & 1.22$\cdot 10^{-1}$ & -0.55& 1.01 & 9.68$\cdot 10^{-4}$ & 0.19& 1.91 & 1.40$\cdot 10^{-1}$ & -0.61 \\ 
F-6 & 6.06 & 1.73 & -0.71& 1.01 & 9.23$\cdot 10^{-4}$ & 0.18& 6.10 & 1.74 & -0.61 \\ 
G-0 & 0.81 & 2.82$\cdot 10^{-2}$ & -0.70& 0.85 & 1.37$\cdot 10^{-2}$ & -0.37& 0.78 & 2.79$\cdot 10^{-2}$ & -0.61 \\ 
G-1 & 0.86 & 3.42$\cdot 10^{-2}$ & -0.56& 0.97 & 1.78$\cdot 10^{-2}$ & -0.37& 0.89 & 3.57$\cdot 10^{-2}$ & -0.62 \\ 
G-2 & 1.26 & 7.91$\cdot 10^{-2}$ & -0.52& 1.26 & 5.15$\cdot 10^{-2}$ & 0.04& 1.26 & 7.22$\cdot 10^{-2}$ & -0.60 \\ 
G-3 & 1.60 & 1.11$\cdot 10^{-1}$ & -0.62& 1.01 & 1.06$\cdot 10^{-3}$ & 0.35& 1.59 & 1.16$\cdot 10^{-1}$ & -0.60 \\ 
G-4 & 2.49 & 2.89$\cdot 10^{-1}$ & -0.68& 1.01 & 9.73$\cdot 10^{-4}$ & 0.19& 2.95 & 1.75$\cdot 10^{-1}$ & -1.29 \\ 
G-5 & 4.93 & 1.22 & -0.65& 1.01 & 1.04$\cdot 10^{-3}$ & 0.19& 2.41 & 2.67$\cdot 10^{-1}$ & -0.63 \\ 
G-6 & 1.58 & 1.13$\cdot 10^{-1}$ & -0.64& 1.01 & 1.10$\cdot 10^{-3}$ & 0.37& 1.59 & 1.16$\cdot 10^{-1}$ & -0.60 \\ 
H-0 & 0.82 & 2.92$\cdot 10^{-2}$ & -0.64& 0.85 & 1.40$\cdot 10^{-2}$ & -0.37& 0.78 & 2.85$\cdot 10^{-2}$ & -0.59 \\ 
H-1 & 0.87 & 3.55$\cdot 10^{-2}$ & -0.64& 0.97 & 1.80$\cdot 10^{-2}$ & -0.37& 0.89 & 3.65$\cdot 10^{-2}$ & -0.63 \\ 
H-2 & 1.31 & 7.27$\cdot 10^{-2}$ & -0.69& 1.26 & 5.16$\cdot 10^{-2}$ & 0.01& 1.26 & 7.21$\cdot 10^{-2}$ & -0.60 \\ 
H-3 & 1.69 & 1.29$\cdot 10^{-1}$ & -0.65& 1.02 & 1.27$\cdot 10^{-3}$ & 0.50& 1.59 & 1.18$\cdot 10^{-1}$ & -0.63 \\ 
H-4 & 4.62 & 9.79$\cdot 10^{-1}$ & -0.61& 1.01 & 9.36$\cdot 10^{-4}$ & 0.19& 2.43 & 2.51$\cdot 10^{-1}$ & -0.52 \\ 
H-5 & 2.46 & 2.82$\cdot 10^{-1}$ & -0.74& 1.01 & 9.01$\cdot 10^{-4}$ & 0.18& 2.95 & 1.88$\cdot 10^{-1}$ & -1.25 \\ 
H-6 & 1.52 & 1.12$\cdot 10^{-1}$ & -0.58& 1.01 & 1.03$\cdot 10^{-3}$ & 0.37& 1.58 & 1.18$\cdot 10^{-1}$ & -0.62 \\ 
I-0 & 0.78 & 2.66$\cdot 10^{-2}$ & -0.67& 0.86 & 1.38$\cdot 10^{-2}$ & -0.61& 0.78 & 2.82$\cdot 10^{-2}$ & -0.60 \\ 
I-1 & 0.94 & 4.00$\cdot 10^{-2}$ & -0.64& 0.99 & 1.95$\cdot 10^{-2}$ & -0.60& 0.90 & 3.78$\cdot 10^{-2}$ & -0.61 \\ 
I-2 & 1.51 & 9.41$\cdot 10^{-2}$ & -0.62& 1.39 & 1.10$\cdot 10^{-1}$ & 0.24& 1.54 & 1.07$\cdot 10^{-1}$ & -0.62 \\ 
I-3 & 5.89 & 1.71 & -0.61& 1.01 & 9.07$\cdot 10^{-4}$ & 0.18& 6.03 & 1.63 & -0.68 \\ 
I-4 & 1.57 & 1.07$\cdot 10^{-1}$ & -0.69& 1.01 & 8.52$\cdot 10^{-4}$ & 0.18& 1.73 & 9.84$\cdot 10^{-2}$ & -0.63 \\ 
I-5 & 2.26 & 2.26$\cdot 10^{-1}$ & -0.67& 1.01 & 1.02$\cdot 10^{-3}$ & 0.21& 1.62 & 1.19$\cdot 10^{-1}$ & -0.60 \\ 
I-6 & 6.29 & 1.84 & -0.59& 1.01 & 1.07$\cdot 10^{-3}$ & 0.20& 5.99 & 1.68 & -0.59 
\end{tabular}
\end{table*}
\endgroup

\subsection{Results for sets A to E [low $\epsilon_d$]}

\begin{figure}
    \centering
    \includegraphics[scale=1]{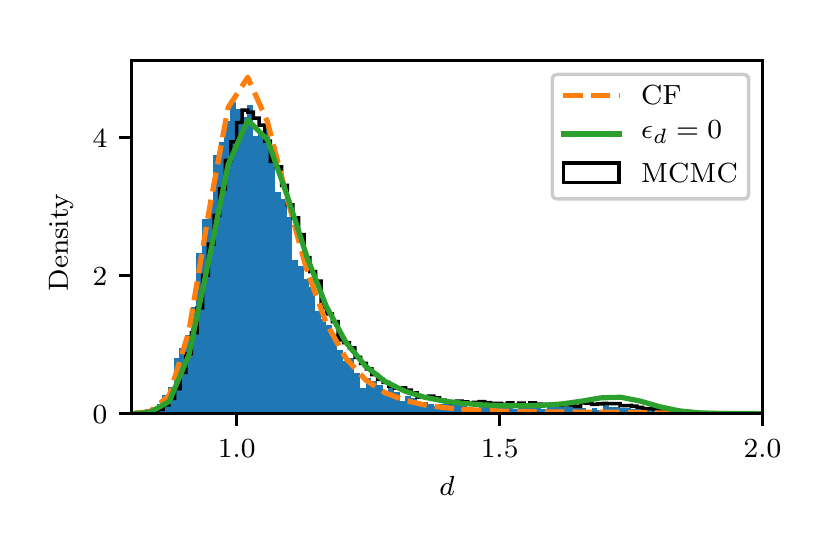}
    \caption{Posterior probability density of the normalized luminosity distance $d$ recovered by LALInference  using a $d^2$ prior (histogram), compared with different approximations indicated with lines obtained with simulation A-2 (see Tables~\ref{tab:seriesparameter} and \ref{tab:series_ext}). Orange dashed line: corrected CF approximation in Eq.~\eqref{eq:CF-again1}; green solid line: analytical solution for $\epsilon_d=0$ in Eq.~\eqref{eq:bessel_posterior}; black step line: MCMC marginalization of Eq.~\eqref{eq:longlong}. }
    \label{fig:BAcase}
\end{figure}

We find that the CF approximation reproduces the LALInference results when $\epsilon_d \lesssim 0.8$. An example is given in Fig.~\ref{fig:BAcase} which shows a comparison of the posterior obtained by LALInference with our predictions in the case of $\epsilon_d=0.03$. In this case, the mean of normalized luminosity distance is approximately $1$.

In the range of low $\epsilon_d$, all methods (MCMC marginalization, CF approximation and the approximation obtained specifically for $\epsilon_d=0$) give consistent results. We conclude that, for $\epsilon_d \lesssim 0.8$, the CF approximation provides a good proxy for the luminosity distance uncertainties. 

The skewnesses reported in Table~\ref{eq:full_table} shows that the shape of the posterior on $d$ can significantly deviates that of a standard Gaussian and symmetrical curve, including in case where the almost totally of the SNR is recovered.

For GW signals with $v>0.5$ the posterior has tails towards smaller luminosity distances, while for $v<0.5$ it has tails towards higher luminosity distances. These tails plays a role in the estimation uncertainties on $H_0$. 

\subsection{Results for sets F to I [high $\epsilon_d$]}

\begin{figure}
    \centering
    \includegraphics[scale=1]{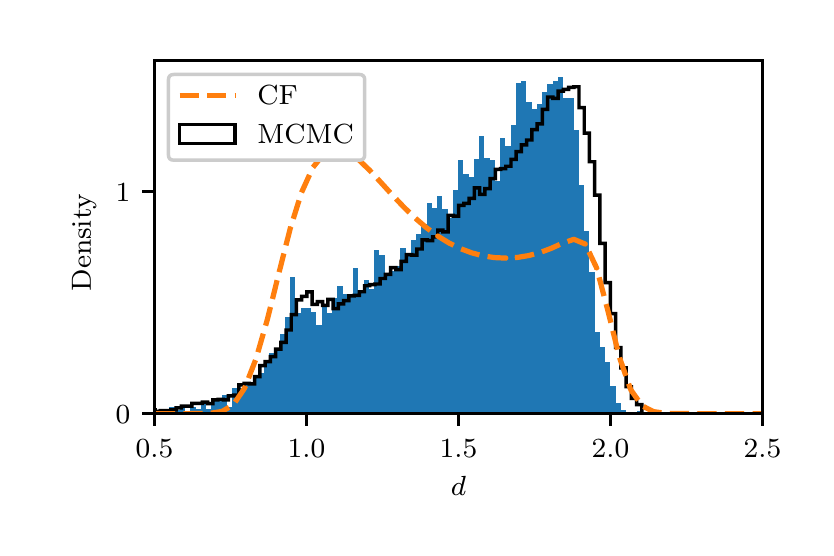}
    \caption{Posterior probability density of the normalized luminosity distance $d$ recovered by LALInference using a $d^2$ prior  (histogram), compared with different approximations indicated with lines obtained with simulation I-2 (see Tables~\ref{tab:seriesparameter} and \ref{tab:series_ext}). Orange dashed line: corrected CF approximation in Eq.~\eqref{eq:CF-again1}, black step line: MCMC marginalization of Eq.~\eqref{eq:longlong}. The CF approximation breaks down  when $\epsilon_d \approx 1$ due to the fact that the maximum likelihood in strongly degenerate in the extrinsic parameter space.}
    \label{fig:BIcase}
\end{figure}

Fig.~\ref{fig:BIcase} compares the results obtained with LALInference, the CF approximant and the MCMC marginalization of Eq.~\eqref{eq:longlong} in a case where $\epsilon_d\gtrsim 0.8$.
The CF approximation is clearly no longer valid, while the MCMC marginalization method still provides a correct approximation of the posterior distribution.

Our explanation is that, in the range $\epsilon_d \sim 1$, the posterior is mostly determined by the prior distribution and only weakly by the data. The prior has a crucial impact on the posterior probability density function when the likelihood is degenerate with respect to the extrinsic parameters, namely when $\epsilon_d \sim 1$. In this case, the prior acts as a selection criterion during the sampling of the degenerate peaks. For instance face-on binaries at higher distances will be arbitrarily preferred if a standard distance prior in $d^2$ is used. This thus leads to a bias that propagates to the estimation of the Hubble constant (discussed in the next Section). Hence, we conclude that care needs to be taken when studying events with $\epsilon_d \sim 1$. No such effect occurs where $\epsilon_d \ll 1$ as the likelihood has no degeneracies in the parameter space.

\begin{figure}
    \centering
    \includegraphics{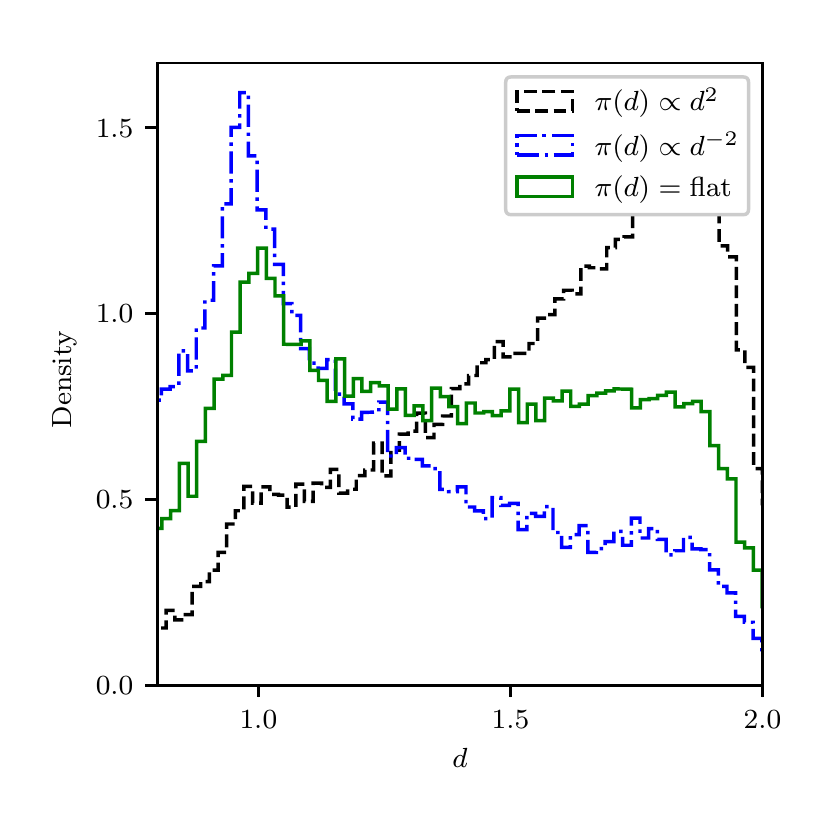}
    \caption{Histogram of the posterior probability density of the normalized luminosity distance for the simulation I-2. The posteriors are obtained with the MCMC marginalization of Eq.~\eqref{eq:longlong} and with different priors on the luminosity distance.}
    \label{fig:efpBI}
\end{figure}

\begin{figure}
    \centering
    \includegraphics{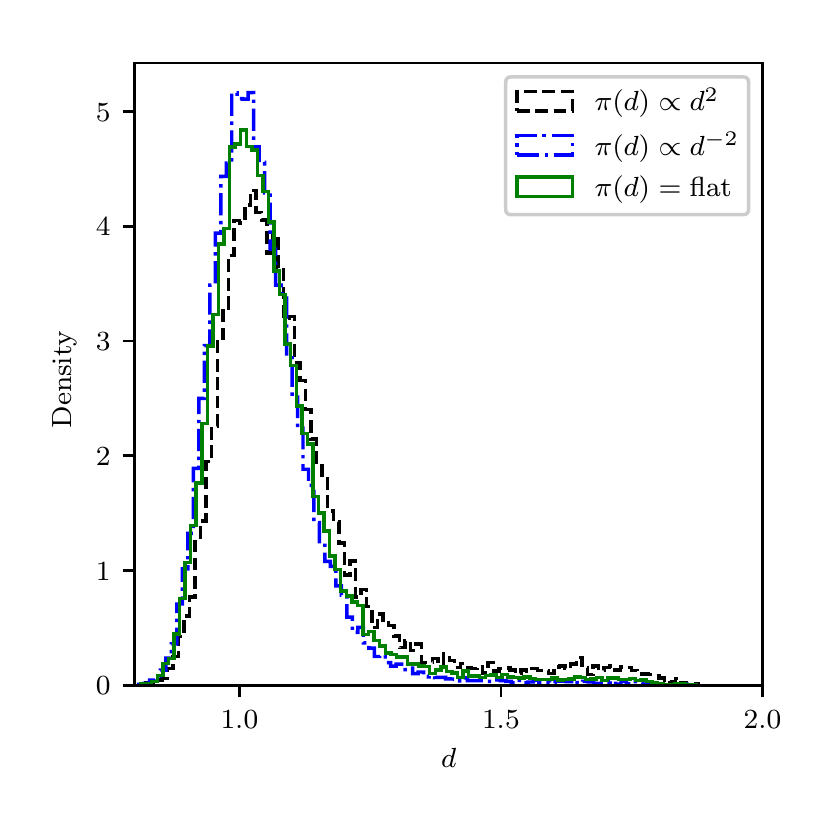}
    \caption{Histogram of the posterior probability density function on the normalized luminosity distance for the injection A-2. The posteriors have been obtained from the MCMC marginalization of Eq.~\eqref{eq:longlong} and changing the prior on the luminosity distance.}
    \label{fig:efPBA}
\end{figure}

Simulation I-2 with $\epsilon_d=1$ provides a good example as the final shape of the posterior probability density distribution that is essentially decided by the prior distribution which scales as $d^2$
We demonstrate this interpretation in Fig.~\ref{fig:efpBI} and Fig.~\ref{fig:efPBA} which show the posterior of the normalized luminosity distance obtained with the MCMC marginalization of Eq.~\eqref{eq:longlong} using different priors. The posterior clearly changes with the prior profile in Fig.~\ref{fig:efpBI} where $\epsilon_d \sim 1$
while the posterior changes marginally in Fig.~\ref{fig:efPBA} where $\epsilon_d=0.03$.  

\section{Application to GW170817 and implications for the Hubble constant estimation \label{sec:6}}

We now apply the approximants derived in Sec.~\ref{sec:4} to the case of GW170817 and draw the implication on the $H_0$ estimation from that source. 
We consider the following event properties: the detection epoch is $t_{\rm gps}=1187008882.43 \, s$ (Aug 17 2017 12:41:04.43 UTC) \cite{Abbott:2018wiz}; the source sky location is that of NGC 4993, i.e., R.A.=$197.45~\rm{deg}$, dec$=-23.38~\rm{deg}$ \cite{2017ApJ...848L..12A} and the detected SNR is  $32.4$. From this information, we obtain $\epsilon_d \approx 0.80$ for the three-detector HLV network. This case is not symptomatic of degeneracies in the likelihood as shown by the set of injections E in the previous section.

In the following simulations, we have also fixed the distance and inclination of the source to the symmetric interval of the parameters reported in \cite{ligo2017gravitational},  $d_{\rm BNS}=41.1^{+4.0}_{-7.3}$~Mpc and $\iota_{\rm{BNS}}=152^{+14}_{-17}$ deg. The luminosity distance posterior is obtained with the CF approximation in Eq.~(\ref{eq:CF-again1}) and using MCMC marginalization of Eq.~\eqref{eq:longlong}. Hence we estimate the symmetric intervals from the simulated posteriors, obtaining  $d_{\rm CF}=40.9^{+4.6}_{-6.8}$~Mpc for the CF approximant and $d_{\rm MCMC}=40.8^{+4.6}_{-7.0}$~Mpc for the MCMC posterior integration.

\begin{figure}
    \centering
    \includegraphics{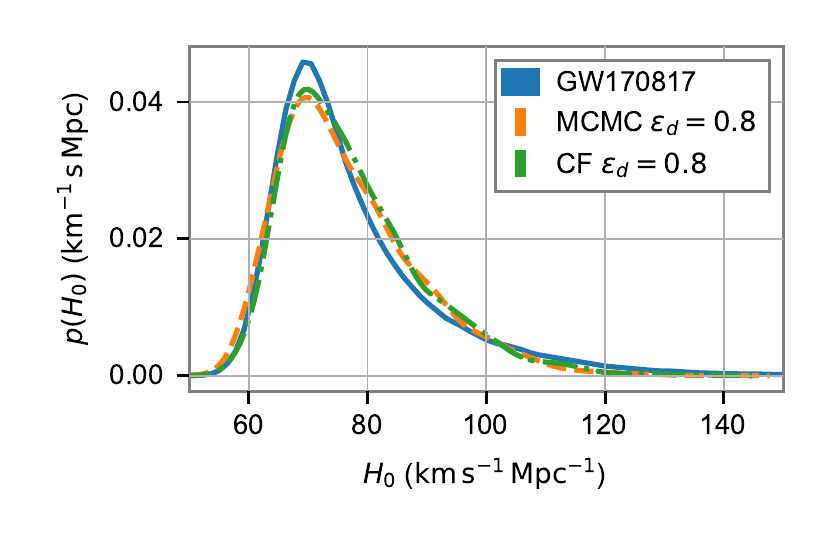}
    \caption{Posterior density of the $H_0$ measure from GW170817 (blue line) using posterior samples available at \url{https://dcc.ligo.org/LIGO-P1700296/public} compared to MCMC marginalization of Eq.~\eqref{eq:longlong} (orange line) and CF approximant (green line). }
    \label{fig:H-comp}
\end{figure}

Using the statistical model in \cite{ligo2017gravitational} and the reported recessional and peculiar velocities of NGC 4993 \cite{ligo2017gravitational}, we determine the posterior of $H_0$ using the luminosity distance predictions obtained above. The results are shown in Fig.~\ref{fig:H-comp} and compared to \cite{ligo2017gravitational}. The predictions fit well the observations. The Hubble constant estimation for GW170817 is $H_0=70.0^{+12.0}_{-8.0} \;\SI{}{km\; s^{-1}\;Mpc^{-1}}$ \cite{ligo2017gravitational}, while our predicted value is $H_0=69.9^{+13.0}_{-7.5} \;\SI{}{km\; s^{-1}\;Mpc^{-1}}$ at $1\sigma$ confidence level by sampling from Eq.~\eqref{eq:longlong} and $H_0=69.9^{+11.2}_{-8.7} \;\SI{}{km\; s^{-1}\;Mpc^{-1}}$ using the corrected CF approximation. The confidence level intervals are comparable with the ones obtained from GW170817 observation.

We simulated the $H_0$ measurement from a ``GW170817-like'' event assuming a source sky location where $\epsilon_d=0$. For this simulation we used the parameters of GW170817 except for the sky location. We used the CF approximant and Eq.~\eqref{eq:longlong} to obtain the Hubble constant posterior. 

\begin{figure}
    \centering
    \includegraphics{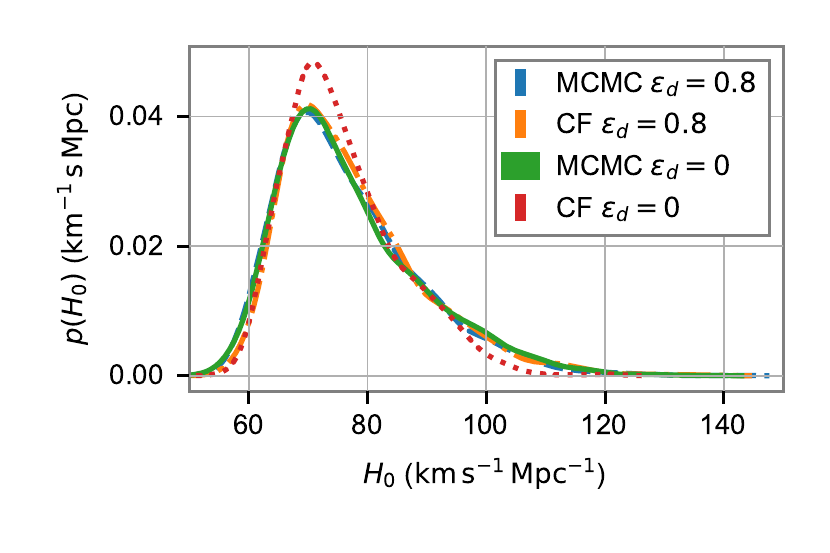}
    \caption{Posterior density of the $H_0$ measure from a GW170817-like event assuming a source sky location where $\epsilon_d=0$ (green and red lines) and where $\epsilon_d \approx 1$ (blue and orange), similarly to Fig.~\ref{fig:H-comp}.}
    \label{fig:H-0sim}
\end{figure}

The resulting $H_0$ posterior in Fig.~\ref{fig:H-0sim} should be compared with Fig.~\ref{fig:H-comp}. The CF approximation for $\epsilon_d=0$ predicts $H_0=70.1^{+10.3}_{-7.0} \;\SI{}{km\; s^{-1}\;Mpc^{-1}}$ while the MCMC marginalization of Eq.~\eqref{eq:longlong} predicts $H_0=69.2^{+13.3}_{-6.6} \;\SI{}{km\; s^{-1}\;Mpc^{-1}}$.
The measurement of the Hubble constant is thus 15\% more accurate if the source sky position has $\epsilon_d\approx 0$ for the CF approximation compared to the actual source location with $\epsilon_d\approx 0.8$. The accuracy obtained with the MCMC marginalization is the same in both the $\epsilon_d \approx 0$  and $\epsilon_d \approx 0.8$ cases.

\section{Conclusions \label{sec:conclusions}}

In this paper, starting from the framework introduced in \cite{cutler1994gravitational}, we have developed several analytic approximants of the posterior of the luminosity distance for localized binary neutron star events. 

To do so, we initially demonstrated the important role of the detector-network and sky-position dependent parameter $\epsilon_d$, which quantifies the ability of the detector network to disentangle the extrinsic parameters encoded in the gravitational-wave polarizations. Then we studied the degeneracies of the likelihood for GW events in absence of noise. We showed that for $\epsilon_d < 1$, the likelihood admits one global maximum which corresponds to the true parameters of the event, while for $\epsilon_d = 1$, the likelihood admits many degenerate global maxima, as it is not possible to disentangle the two gravitational-wave polarizations.

We obtained different approximants of the likelihood from Eq.~\eqref{eq:longlong}. One of these approximants was already presented in \cite{cutler1994gravitational}, though with a small (but significant) error.  The error on the extrinsic parameter estimation predicted by the approximants were compared with those obtained for 63 simulated gravitational-wave signals using a Bayesian sampler available in the LALInference software package. This comparison confirms that, for sky locations with $\epsilon_d\sim 1$, the likelihood is degenerate, and this can lead to a prior-induced bias in the distance estimate.

We verified that GW170817 is localized in a sky patch with $\epsilon_d \approx 0.80$, meaning that we do not expect the likelihood for this GW event to be uninformative. We have checked that the posterior on the luminosity distance obtained with our approximants is compatible with the one reported in \cite{Abbott:2018wiz}. We checked that the Hubble constant estimates based on the statistical model detailed in \cite{ligo2017gravitational} and obtained using both our approximations and the published posterior samples are consistent.

The new approximants we have developed can be employed to forecast and study the accuracy of the $H_0$ measurement in different observing scenarios. For instance they can be employed in a framework similar to \cite{2018arXiv181111723M} in order to evaluate scenarios with different detector networks and duty cycles. Moreover, the $\epsilon_d$ parameter provides a simple and quick indicator to understand whether the luminosity distance posterior is informative or not. This parameter can also be used to evaluate or optimize the geometry and position of an additional detector to reduce the regions in the sky where the likelihood is degenerate.

\section*{Acknowledgements}

S.~M. acknowledges the UnivEarthS Labex program (ANR-10-LABX-0023) for financial support. K.~L. and D.~S. are grateful to CERN and the University of Geneva (D.S) for hospitality whilst this work was in progress. We thank C.-J.~Haster, K.~Chatziioannou, S.~Vitale, W.~del Pozzo, V.~Raymond, B.~Malte Sch\"afer, E.~Porter, M.~Barsuglia, I.~Magana and H.-Y.~Chen for useful discussions and for helping with the LALInference runs. We are grateful to the LIGO Scientific Collaboration for sharing the LAL software suite \cite{lalsuite}. The authors are also grateful for the computational resources provided by the LIGO Laboratory. 

\appendix

\section{Degeneracy of the maximum likelihood template}
\label{AppA}

\subsection{Case of $\epsilon_d < 1$}

We define two templates to be degenerate with respect to a GW signal if they both maximize the likelihood in Eq.~\ref{eq:likelihood}, or equivalently Eq.~\eqref{eq:liklik}.

When $\epsilon_d > 0$, Eq.~\eqref{eq:liklik} is maximized when $\bar{\mathcal{A}}_{+/\times} = \bar{\mathcal{A}}'_{+/\times}$. Writing the two effective polarizations in terms of the extrinsic parameters $D, \psi, v$ yields the two equations (for simplicity of notation, we reabsorb the phases $\phi_c$ and $\phi'_c$ into a redefinition of $\bar{\psi}$ and  $\bar{\psi}'$ respectively)
\begin{align}
    & (d\chi_+ \cos 2 \bar{\psi} -\chi'_+ \cos 2 \bar{\psi}'  )^2+(d v \sin 2 \bar{\psi} -v' \sin 2 \bar{\psi}'  )^2 =0 
    \nonumber \\
    & (d\chi_+ \sin 2 \bar{\psi}-\chi'_+ \sin 2 \bar{\psi}' )^2+(d v \cos 2 \bar{\psi} -v' \cos 2 \bar{\psi}' )^2 =0,
    \nonumber
\end{align}
whose sum gives
\begin{equation}
    d^2(\chi^2_++v^2)-2d\cos(2\psi_-)(\chi_+\chi'_+ + vv')+\chi_+'^2+v'^2=0.
    \nonumber
\end{equation}

The two solutions of this equation are
\begin{equation}
    d_\pm=\frac{\cos 2\psi_- (\chi_+ \chi'_+ + vv') \pm \sqrt{\Delta}}{\chi_+^2+v^2},
     \nonumber
\end{equation}
where 
\begin{equation}
    \Delta= (\cos^2 \psi_- -1)(\chi_+ \chi_+' +vv')^2-(\chi_+ v'-\chi_+'v)^2.
     \label{eq:nearly}
\end{equation}

Note that if $\Delta > 0$ then possibly a degenerate template with $d \neq 1$ would exist, while if $\Delta<0$  then no degenerate solutions exist. Clearly in the present case, it follows from Eq.~\eqref{eq:nearly} that $\Delta$ is negative unless $\psi_-=0$ and consequently $v'=v$. The other solution $v'=1/v$ is permitted only for $v=1$ since $(|v|,|v'|) \leq 1$. Hence we obtain the solution $d_{\pm}=1$, which is the correct luminosity distance. 

To conclude, (at the 0 PN order and for $\epsilon_d < 1$) no degenerate templates which are able to maximize the likelihood exist in the parameter space. Furthermore, if noise terms are neglected, the only template which is able to maximize the likelihood is the one corresponding to the GW signal present.

\subsection{Case of $\epsilon_d \approx 1$}

Let us assume that $\epsilon_d=1$, so that the term in $\bar{\mathcal{A}}_\times$ vanishes in the likelihood (see Eq.~\eqref{eq:liklik}). 
%the only condition for its maximization is that $\bar{\mathcal{A}}_{+} = \bar{\mathcal{A}}'_{+}$. 
Hence the likelihood is maximised when $\bar{\mathcal{A}}_{+} = \bar{\mathcal{A}}'_{+}$ leading to two equations (from the real and imaginary parts)
\footnote{For simplicity, we have reabsorbed the phases $\phi_c$ and $\phi'_c$ into a redefinition of $\bar{\psi}$ and  $\bar{\psi}'$ respectively.}:
\begin{align}
    \chi_+' \cos 2 \bar{\psi}' &=d \chi_+ \cos 2 \bar{\psi} \label{eq:1sys}\\
    v' \sin 2\bar{ \psi'} &=d v \sin 2 \bar{\psi},
    \label{eq:2sys}
\end{align}
On taking the sum of the square of the two previous equations, we obtain a solution for $d$:
 \begin{equation}
     d =\bigg[ \bigg(\frac{\chi_+}{\chi_+'}\bigg)^2 \cos^2 2\bar{\psi} + \bigg(\frac{v}{v'}\bigg)^2 \sin^2 2\bar{\psi}  \bigg]^{-1/2}.
     \label{eq:dsolution}
 \end{equation}
 Since Eq.~\eqref{eq:dsolution} is positive definite, it represents a physical distance. However, to prove that Eq.~\eqref{eq:dsolution} represents a physical template, we substitute back into one of the two Eqs.~\eqref{eq:1sys}-\eqref{eq:2sys} and obtain
\begin{equation}
\sin 2 \bar{\psi}'=\bigg[1+\bigg(\frac{v'}{v}\frac{\chi_+}{\chi_+'} \cot 2 \bar{\psi} \bigg)^2\bigg]^{-1/2}.    
\label{eq:psideg}
\end{equation}
Since the RHS in Eq.~\eqref{eq:psideg} is $<1$, there are degenerate templates that maximize the likelihood. 

\noindent
\bibliography{refs}

\begin{thebibliography}{39}%
\makeatletter
\providecommand \@ifxundefined [1]{%
 \@ifx{#1\undefined}
}%
\providecommand \@ifnum [1]{%
 \ifnum #1\expandafter \@firstoftwo
 \else \expandafter \@secondoftwo
 \fi
}%
\providecommand \@ifx [1]{%
 \ifx #1\expandafter \@firstoftwo
 \else \expandafter \@secondoftwo
 \fi
}%
\providecommand \natexlab [1]{#1}%
\providecommand \enquote  [1]{``#1''}%
\providecommand \bibnamefont  [1]{#1}%
\providecommand \bibfnamefont [1]{#1}%
\providecommand \citenamefont [1]{#1}%
\providecommand \href@noop [0]{\@secondoftwo}%
\providecommand \href [0]{\begingroup \@sanitize@url \@href}%
\providecommand \@href[1]{\@@startlink{#1}\@@href}%
\providecommand \@@href[1]{\endgroup#1\@@endlink}%
\providecommand \@sanitize@url [0]{\catcode `\\12\catcode `\$12\catcode
  `\&12\catcode `\#12\catcode `\^12\catcode `\_12\catcode `\%12\relax}%
\providecommand \@@startlink[1]{}%
\providecommand \@@endlink[0]{}%
\providecommand \url  [0]{\begingroup\@sanitize@url \@url }%
\providecommand \@url [1]{\endgroup\@href {#1}{\urlprefix }}%
\providecommand \urlprefix  [0]{URL }%
\providecommand \Eprint [0]{\href }%
\providecommand \doibase [0]{http://dx.doi.org/}%
\providecommand \selectlanguage [0]{\@gobble}%
\providecommand \bibinfo  [0]{\@secondoftwo}%
\providecommand \bibfield  [0]{\@secondoftwo}%
\providecommand \translation [1]{[#1]}%
\providecommand \BibitemOpen [0]{}%
\providecommand \bibitemStop [0]{}%
\providecommand \bibitemNoStop [0]{.\EOS\space}%
\providecommand \EOS [0]{\spacefactor3000\relax}%
\providecommand \BibitemShut  [1]{\csname bibitem#1\endcsname}%
\let\auto@bib@innerbib\@empty
%</preamble>
\bibitem [{\citenamefont {Abbott}\ \emph {et~al.}(2016)\citenamefont {Abbott}
  \emph {et~al.}}]{Abbott:2016blz}%
  \BibitemOpen
  \bibfield  {author} {\bibinfo {author} {\bibfnamefont {B.~P.}\ \bibnamefont
  {Abbott}} \emph {et~al.} (\bibinfo {collaboration} {LIGO Scientific
  Collaboration, Virgo}),\ }\href {\doibase 10.1103/PhysRevLett.116.061102}
  {\bibfield  {journal} {\bibinfo  {journal} {Phys.~Rev.~Lett.}\ }\textbf
  {\bibinfo {volume} {116}},\ \bibinfo {pages} {061102} (\bibinfo {year}
  {2016})},\ \Eprint {http://arxiv.org/abs/1602.03837} {arXiv:1602.03837
  [gr-qc]} \BibitemShut {NoStop}%
%%CITATION = ARXIV:1602.03837;%%
\bibitem [{\citenamefont {Abbott}\ \emph {et~al.}(2018)\citenamefont {Abbott}
  \emph {et~al.}}]{LIGOScientific:2018mvr}%
  \BibitemOpen
  \bibfield  {author} {\bibinfo {author} {\bibfnamefont {B.~P.}\ \bibnamefont
  {Abbott}} \emph {et~al.} (\bibinfo {collaboration} {LIGO Scientific
  Collaboration, Virgo}),\ }\href@noop {} {\  (\bibinfo {year} {2018})},\
  \Eprint {http://arxiv.org/abs/1811.12907} {arXiv:1811.12907 [astro-ph.HE]}
  \BibitemShut {NoStop}%
%%CITATION = ARXIV:1811.12907;%%
\bibitem [{\citenamefont {Abbott}\ \emph
  {et~al.}(2019{\natexlab{a}})\citenamefont {Abbott} \emph
  {et~al.}}]{LIGOScientific:2019fpa}%
  \BibitemOpen
  \bibfield  {author} {\bibinfo {author} {\bibfnamefont {B.~P.}\ \bibnamefont
  {Abbott}} \emph {et~al.} (\bibinfo {collaboration} {LIGO Scientific
  Collaboration, Virgo}),\ }\href@noop {} {\  (\bibinfo {year}
  {2019}{\natexlab{a}})},\ \Eprint {http://arxiv.org/abs/1903.04467}
  {arXiv:1903.04467 [gr-qc]} \BibitemShut {NoStop}%
%%CITATION = ARXIV:1903.04467;%%
\bibitem [{\citenamefont {Abbott}\ \emph
  {et~al.}(2017{\natexlab{a}})\citenamefont {Abbott} \emph
  {et~al.}}]{TheLIGOScientific:2017qsa}%
  \BibitemOpen
  \bibfield  {author} {\bibinfo {author} {\bibfnamefont {B.~P.}\ \bibnamefont
  {Abbott}} \emph {et~al.} (\bibinfo {collaboration} {LIGO Scientific
  Collaboration, Virgo}),\ }\href {\doibase 10.1103/PhysRevLett.119.161101}
  {\bibfield  {journal} {\bibinfo  {journal} {Phys.~Rev.~Lett.}\ }\textbf
  {\bibinfo {volume} {119}},\ \bibinfo {pages} {161101} (\bibinfo {year}
  {2017}{\natexlab{a}})},\ \Eprint {http://arxiv.org/abs/1710.05832}
  {arXiv:1710.05832 [gr-qc]} \BibitemShut {NoStop}%
%%CITATION = ARXIV:1710.05832;%%
\bibitem [{\citenamefont {{Somiya}}(2012)}]{2012CQGra..29l4007S}%
  \BibitemOpen
  \bibfield  {author} {\bibinfo {author} {\bibfnamefont {K.}~\bibnamefont
  {{Somiya}}},\ }\href {\doibase 10.1088/0264-9381/29/12/124007} {\bibfield
  {journal} {\bibinfo  {journal} {Class.~Quantum~Grav.}\ }\textbf {\bibinfo
  {volume} {29}},\ \bibinfo {eid} {124007} (\bibinfo {year} {2012})},\ \Eprint
  {http://arxiv.org/abs/1111.7185} {arXiv:1111.7185 [gr-qc]} \BibitemShut
  {NoStop}%
\bibitem [{\citenamefont {{Unnikrishnan}}(2013)}]{2013IJMPD..2241010U}%
  \BibitemOpen
  \bibfield  {author} {\bibinfo {author} {\bibfnamefont {C.~S.}\ \bibnamefont
  {{Unnikrishnan}}},\ }\href {\doibase 10.1142/S0218271813410101} {\bibfield
  {journal} {\bibinfo  {journal} {Int.~J.~Mod.~Phys.~D}\ }\textbf {\bibinfo
  {volume} {22}},\ \bibinfo {eid} {1341010} (\bibinfo {year} {2013})},\ \Eprint
  {http://arxiv.org/abs/1510.06059} {arXiv:1510.06059 [physics.ins-det]}
  \BibitemShut {NoStop}%
\bibitem [{\citenamefont {{Del Pozzo}}(2012)}]{2012PhRvD..86d3011D}%
  \BibitemOpen
  \bibfield  {author} {\bibinfo {author} {\bibfnamefont {W.}~\bibnamefont {{Del
  Pozzo}}},\ }\href {\doibase 10.1103/PhysRevD.86.043011} {\bibfield  {journal}
  {\bibinfo  {journal} {Phys.~Rev.}\ }\textbf {\bibinfo {volume} {D86}},\
  \bibinfo {eid} {043011} (\bibinfo {year} {2012})},\ \Eprint
  {http://arxiv.org/abs/1108.1317} {arXiv:1108.1317} \BibitemShut {NoStop}%
\bibitem [{\citenamefont {Schutz}(1986)}]{Schutz:1986gp}%
  \BibitemOpen
  \bibfield  {author} {\bibinfo {author} {\bibfnamefont {B.~F.}\ \bibnamefont
  {Schutz}},\ }\href {\doibase 10.1038/323310a0} {\bibfield  {journal}
  {\bibinfo  {journal} {Nature}\ }\textbf {\bibinfo {volume} {323}},\ \bibinfo
  {pages} {310} (\bibinfo {year} {1986})}\BibitemShut {NoStop}%
%%CITATION = NATUA,323,310;%%
\bibitem [{\citenamefont {Jackson}(2015)}]{Jackson2015}%
  \BibitemOpen
  \bibfield  {author} {\bibinfo {author} {\bibfnamefont {N.}~\bibnamefont
  {Jackson}},\ }\href {\doibase 10.1007/lrr-2015-2} {\bibfield  {journal}
  {\bibinfo  {journal} {Living~Rev.~Rel.}\ }\textbf {\bibinfo {volume} {18}},\
  \bibinfo {pages} {2} (\bibinfo {year} {2015})}\BibitemShut {NoStop}%
\bibitem [{\citenamefont {Riess}\ \emph {et~al.}(2016)\citenamefont {Riess},
  \citenamefont {Macri}, \citenamefont {Hoffmann} \emph
  {et~al.}}]{0004-637X-826-1-56}%
  \BibitemOpen
  \bibfield  {author} {\bibinfo {author} {\bibfnamefont {A.~G.}\ \bibnamefont
  {Riess}}, \bibinfo {author} {\bibfnamefont {L.}~\bibnamefont {Macri}},
  \bibinfo {author} {\bibfnamefont {S.~L.}\ \bibnamefont {Hoffmann}},  \emph
  {et~al.},\ }\href {http://stacks.iop.org/0004-637X/826/i=1/a=56} {\bibfield
  {journal} {\bibinfo  {journal} {ApJ}\ }\textbf {\bibinfo {volume} {826}},\
  \bibinfo {pages} {56} (\bibinfo {year} {2016})}\BibitemShut {NoStop}%
\bibitem [{\citenamefont {Ade}\ \emph {et~al.}(2016)\citenamefont {Ade} \emph
  {et~al.}}]{ade2016planck}%
  \BibitemOpen
  \bibfield  {author} {\bibinfo {author} {\bibfnamefont {P.~A.~R.}\
  \bibnamefont {Ade}} \emph {et~al.} (\bibinfo {collaboration} {Planck}),\
  }\href {\doibase 10.1051/0004-6361/201525830} {\bibfield  {journal} {\bibinfo
   {journal} {Astron. Astrophys.}\ }\textbf {\bibinfo {volume} {594}},\
  \bibinfo {pages} {A13} (\bibinfo {year} {2016})},\ \Eprint
  {http://arxiv.org/abs/1502.01589} {arXiv:1502.01589 [astro-ph.CO]}
  \BibitemShut {NoStop}%
%%CITATION = ARXIV:1502.01589;%%
\bibitem [{\citenamefont {{Ishak}}(2019)}]{2019LRR....22....1I}%
  \BibitemOpen
  \bibfield  {author} {\bibinfo {author} {\bibfnamefont {M.}~\bibnamefont
  {{Ishak}}},\ }\href {\doibase 10.1007/s41114-018-0017-4} {\bibfield
  {journal} {\bibinfo  {journal} {Living~Rev.~Rel.}\ }\textbf {\bibinfo
  {volume} {22}},\ \bibinfo {eid} {1} (\bibinfo {year} {2019})},\ \Eprint
  {http://arxiv.org/abs/1806.10122} {arXiv:1806.10122 [astro-ph.CO]}
  \BibitemShut {NoStop}%
\bibitem [{\citenamefont {{The LIGO Scientific Collaboration}}\ \emph
  {et~al.}(2019)\citenamefont {{The LIGO Scientific Collaboration}},
  \citenamefont {{the Virgo Collaboration}},\ and\ \citenamefont
  {{Abbott}}}]{2019arXiv190806060T}%
  \BibitemOpen
  \bibfield  {author} {\bibinfo {author} {\bibnamefont {{The LIGO Scientific
  Collaboration}}}, \bibinfo {author} {\bibnamefont {{the Virgo
  Collaboration}}}, \ and\ \bibinfo {author} {\bibfnamefont {B.~P. e.~a.}\
  \bibnamefont {{Abbott}}},\ }\href@noop {} {\bibfield  {journal} {\bibinfo
  {journal} {arXiv e-prints}\ ,\ \bibinfo {eid} {arXiv:1908.06060}} (\bibinfo
  {year} {2019})},\ \Eprint {http://arxiv.org/abs/1908.06060} {arXiv:1908.06060
  [astro-ph.CO]} \BibitemShut {NoStop}%
\bibitem [{\citenamefont {Abbott}\ \emph
  {et~al.}(2017{\natexlab{b}})\citenamefont {Abbott} \emph
  {et~al.}}]{ligo2017gravitational}%
  \BibitemOpen
  \bibfield  {author} {\bibinfo {author} {\bibfnamefont {B.~P.}\ \bibnamefont
  {Abbott}} \emph {et~al.} (\bibinfo {collaboration} {LIGO Scientific
  Collaboration, VINROUGE, Las Cumbres Observatory, DES, DLT40, Virgo, 1M2H,
  Dark Energy Camera GW-E, MASTER}),\ }\href {\doibase 10.1038/nature24471}
  {\bibfield  {journal} {\bibinfo  {journal} {Nature}\ }\textbf {\bibinfo
  {volume} {551}},\ \bibinfo {pages} {85} (\bibinfo {year}
  {2017}{\natexlab{b}})},\ \Eprint {http://arxiv.org/abs/1710.05835}
  {arXiv:1710.05835 [astro-ph.CO]} \BibitemShut {NoStop}%
%%CITATION = ARXIV:1710.05835;%%
\bibitem [{\citenamefont {Abbott}\ \emph
  {et~al.}(2019{\natexlab{b}})\citenamefont {Abbott} \emph
  {et~al.}}]{Abbott:2018wiz}%
  \BibitemOpen
  \bibfield  {author} {\bibinfo {author} {\bibfnamefont {B.~P.}\ \bibnamefont
  {Abbott}} \emph {et~al.} (\bibinfo {collaboration} {LIGO Scientific
  Collaboration, Virgo}),\ }\href {\doibase 10.1103/PhysRevX.9.011001}
  {\bibfield  {journal} {\bibinfo  {journal} {Phys.~Rev.}\ }\textbf {\bibinfo
  {volume} {X9}},\ \bibinfo {pages} {011001} (\bibinfo {year}
  {2019}{\natexlab{b}})},\ \Eprint {http://arxiv.org/abs/1805.11579}
  {arXiv:1805.11579 [gr-qc]} \BibitemShut {NoStop}%
%%CITATION = ARXIV:1805.11579;%%
\bibitem [{\citenamefont {Abbott}\ \emph
  {et~al.}(2019{\natexlab{c}})\citenamefont {Abbott} \emph
  {et~al.}}]{2019PhRvX...9a1001A}%
  \BibitemOpen
  \bibfield  {author} {\bibinfo {author} {\bibfnamefont {B.~P.}\ \bibnamefont
  {Abbott}} \emph {et~al.} (\bibinfo {collaboration} {LIGO Scientific
  Collaboration, Virgo}),\ }\href {\doibase 10.1103/PhysRevX.9.011001}
  {\bibfield  {journal} {\bibinfo  {journal} {Phys.~Rev.~X}\ }\textbf {\bibinfo
  {volume} {9}},\ \bibinfo {eid} {011001} (\bibinfo {year}
  {2019}{\natexlab{c}})},\ \Eprint {http://arxiv.org/abs/1805.11579}
  {arXiv:1805.11579 [gr-qc]} \BibitemShut {NoStop}%
\bibitem [{\citenamefont {Veitch}\ \emph {et~al.}(2012)\citenamefont {Veitch},
  \citenamefont {Mandel}, \citenamefont {Aylott} \emph
  {et~al.}}]{PhysRevD.85.104045}%
  \BibitemOpen
  \bibfield  {author} {\bibinfo {author} {\bibfnamefont {J.}~\bibnamefont
  {Veitch}}, \bibinfo {author} {\bibfnamefont {I.}~\bibnamefont {Mandel}},
  \bibinfo {author} {\bibfnamefont {B.}~\bibnamefont {Aylott}},  \emph
  {et~al.},\ }\href {\doibase 10.1103/PhysRevD.85.104045} {\bibfield  {journal}
  {\bibinfo  {journal} {Phys.~Rev.}\ }\textbf {\bibinfo {volume} {D85}},\
  \bibinfo {pages} {104045} (\bibinfo {year} {2012})}\BibitemShut {NoStop}%
\bibitem [{\citenamefont {Vallisneri}(2008)}]{vallisneri2008use}%
  \BibitemOpen
  \bibfield  {author} {\bibinfo {author} {\bibfnamefont {M.}~\bibnamefont
  {Vallisneri}},\ }\href@noop {} {\bibfield  {journal} {\bibinfo  {journal}
  {Phys.~Rev.}\ }\textbf {\bibinfo {volume} {D77}},\ \bibinfo {pages} {042001}
  (\bibinfo {year} {2008})}\BibitemShut {NoStop}%
\bibitem [{\citenamefont {Cutler}\ and\ \citenamefont
  {Flanagan}(1994)}]{cutler1994gravitational}%
  \BibitemOpen
  \bibfield  {author} {\bibinfo {author} {\bibfnamefont {C.}~\bibnamefont
  {Cutler}}\ and\ \bibinfo {author} {\bibfnamefont {E.~E.}\ \bibnamefont
  {Flanagan}},\ }\href {\doibase 10.1103/PhysRevD.49.2658} {\bibfield
  {journal} {\bibinfo  {journal} {Phys.~Rev.}\ }\textbf {\bibinfo {volume}
  {D49}},\ \bibinfo {pages} {2658} (\bibinfo {year} {1994})},\ \Eprint
  {http://arxiv.org/abs/gr-qc/9402014} {arXiv:gr-qc/9402014 [gr-qc]}
  \BibitemShut {NoStop}%
%%CITATION = GR-QC/9402014;%%
\bibitem [{\citenamefont {{Veitch}}\ \emph {et~al.}(2015)\citenamefont
  {{Veitch}}, \citenamefont {{Raymond}}, \citenamefont {{Farr}} \emph
  {et~al.}}]{2015PhRvD..91d2003V}%
  \BibitemOpen
  \bibfield  {author} {\bibinfo {author} {\bibfnamefont {J.}~\bibnamefont
  {{Veitch}}}, \bibinfo {author} {\bibfnamefont {V.}~\bibnamefont {{Raymond}}},
  \bibinfo {author} {\bibfnamefont {B.}~\bibnamefont {{Farr}}},  \emph
  {et~al.},\ }\href {\doibase 10.1103/PhysRevD.91.042003} {\bibfield  {journal}
  {\bibinfo  {journal} {Phys.~Rev.}\ }\textbf {\bibinfo {volume} {D91}},\
  \bibinfo {eid} {042003} (\bibinfo {year} {2015})},\ \Eprint
  {http://arxiv.org/abs/1409.7215} {arXiv:1409.7215 [gr-qc]} \BibitemShut
  {NoStop}%
\bibitem [{\citenamefont {Jaranowski}\ and\ \citenamefont
  {Kr{\'o}lak}(2012)}]{Jaranowski2012}%
  \BibitemOpen
  \bibfield  {author} {\bibinfo {author} {\bibfnamefont {P.}~\bibnamefont
  {Jaranowski}}\ and\ \bibinfo {author} {\bibfnamefont {A.}~\bibnamefont
  {Kr{\'o}lak}},\ }\href {\doibase 10.12942/lrr-2012-4} {\bibfield  {journal}
  {\bibinfo  {journal} {Living~Rev.~Rel.}\ }\textbf {\bibinfo {volume} {15}},\
  \bibinfo {pages} {4} (\bibinfo {year} {2012})}\BibitemShut {NoStop}%
\bibitem [{\citenamefont {Maggiore}(2008)}]{maggiore2008gravitational}%
  \BibitemOpen
  \bibfield  {author} {\bibinfo {author} {\bibfnamefont {M.}~\bibnamefont
  {Maggiore}},\ }\href@noop {} {\emph {\bibinfo {title} {Gravitational Waves:
  Volume 1: Theory and Experiments}}},\ Vol.~\bibinfo {volume} {1}\ (\bibinfo
  {publisher} {Oxford university press},\ \bibinfo {year} {2008})\BibitemShut
  {NoStop}%
\bibitem [{\citenamefont {Sathyaprakash}\ and\ \citenamefont
  {Schutz}(2009)}]{Sathyaprakash2009}%
  \BibitemOpen
  \bibfield  {author} {\bibinfo {author} {\bibfnamefont {B.~S.}\ \bibnamefont
  {Sathyaprakash}}\ and\ \bibinfo {author} {\bibfnamefont {B.~F.}\ \bibnamefont
  {Schutz}},\ }\href {\doibase 10.12942/lrr-2009-2} {\bibfield  {journal}
  {\bibinfo  {journal} {Living~Rev.~Rel.}\ }\textbf {\bibinfo {volume} {12}},\
  \bibinfo {pages} {2} (\bibinfo {year} {2009})}\BibitemShut {NoStop}%
\bibitem [{\citenamefont {Finn}\ and\ \citenamefont
  {Chernoff}(1993)}]{Finn:1992xs}%
  \BibitemOpen
  \bibfield  {author} {\bibinfo {author} {\bibfnamefont {L.~S.}\ \bibnamefont
  {Finn}}\ and\ \bibinfo {author} {\bibfnamefont {D.~F.}\ \bibnamefont
  {Chernoff}},\ }\href {\doibase 10.1103/PhysRevD.47.2198} {\bibfield
  {journal} {\bibinfo  {journal} {Phys.~Rev.}\ }\textbf {\bibinfo {volume}
  {D47}},\ \bibinfo {pages} {2198} (\bibinfo {year} {1993})},\ \Eprint
  {http://arxiv.org/abs/gr-qc/9301003} {arXiv:gr-qc/9301003 [gr-qc]}
  \BibitemShut {NoStop}%
%%CITATION = GR-QC/9301003;%%
\bibitem [{\citenamefont {{Jaranowski}}\ \emph {et~al.}(1998)\citenamefont
  {{Jaranowski}}, \citenamefont {{Kr{\'o}lak}},\ and\ \citenamefont
  {{Schutz}}}]{1998PhRvD..58f3001J}%
  \BibitemOpen
  \bibfield  {author} {\bibinfo {author} {\bibfnamefont {P.}~\bibnamefont
  {{Jaranowski}}}, \bibinfo {author} {\bibfnamefont {A.}~\bibnamefont
  {{Kr{\'o}lak}}}, \ and\ \bibinfo {author} {\bibfnamefont {B.~F.}\
  \bibnamefont {{Schutz}}},\ }\href {\doibase 10.1103/PhysRevD.58.063001}
  {\bibfield  {journal} {\bibinfo  {journal} {Phys.~Rev.}\ }\textbf {\bibinfo
  {volume} {D58}},\ \bibinfo {eid} {063001} (\bibinfo {year} {1998})},\ \Eprint
  {http://arxiv.org/abs/gr-qc/9804014} {arXiv:gr-qc/9804014 [gr-qc]}
  \BibitemShut {NoStop}%
\bibitem [{\citenamefont {Markovi\ifmmode~\acute{c}\else
  \'{c}\fi{}}(1993)}]{PhysRevD.48.4738}%
  \BibitemOpen
  \bibfield  {author} {\bibinfo {author} {\bibfnamefont {D.}~\bibnamefont
  {Markovi\ifmmode~\acute{c}\else \'{c}\fi{}}},\ }\href {\doibase
  10.1103/PhysRevD.48.4738} {\bibfield  {journal} {\bibinfo  {journal}
  {Phys.~Rev.}\ }\textbf {\bibinfo {volume} {D48}},\ \bibinfo {pages} {4738}
  (\bibinfo {year} {1993})}\BibitemShut {NoStop}%
\bibitem [{\citenamefont {Jaranowski}\ and\ \citenamefont
  {Krolak}(1994)}]{PhysRevD.49.1723}%
  \BibitemOpen
  \bibfield  {author} {\bibinfo {author} {\bibfnamefont {P.}~\bibnamefont
  {Jaranowski}}\ and\ \bibinfo {author} {\bibfnamefont {A.}~\bibnamefont
  {Krolak}},\ }\href {\doibase 10.1103/PhysRevD.49.1723} {\bibfield  {journal}
  {\bibinfo  {journal} {Phys.~Rev.}\ }\textbf {\bibinfo {volume} {D49}},\
  \bibinfo {pages} {1723} (\bibinfo {year} {1994})}\BibitemShut {NoStop}%
\bibitem [{\citenamefont {{Vitale}}\ and\ \citenamefont
  {{Chen}}(2018)}]{2018PhRvL.121b1303V}%
  \BibitemOpen
  \bibfield  {author} {\bibinfo {author} {\bibfnamefont {S.}~\bibnamefont
  {{Vitale}}}\ and\ \bibinfo {author} {\bibfnamefont {H.-Y.}\ \bibnamefont
  {{Chen}}},\ }\href {\doibase 10.1103/PhysRevLett.121.021303} {\bibfield
  {journal} {\bibinfo  {journal} {Physical Review Letters}\ }\textbf {\bibinfo
  {volume} {121}},\ \bibinfo {eid} {021303} (\bibinfo {year} {2018})},\ \Eprint
  {http://arxiv.org/abs/1804.07337} {arXiv:1804.07337} \BibitemShut {NoStop}%
\bibitem [{\citenamefont {{Chen}}\ \emph {et~al.}(2018)\citenamefont {{Chen}},
  \citenamefont {{Vitale}},\ and\ \citenamefont
  {{Narayan}}}]{2018arXiv180705226C}%
  \BibitemOpen
  \bibfield  {author} {\bibinfo {author} {\bibfnamefont {H.-Y.}\ \bibnamefont
  {{Chen}}}, \bibinfo {author} {\bibfnamefont {S.}~\bibnamefont {{Vitale}}}, \
  and\ \bibinfo {author} {\bibfnamefont {R.}~\bibnamefont {{Narayan}}},\
  }\href@noop {} {\bibfield  {journal} {\bibinfo  {journal} {arXiv e-prints}\
  ,\ \bibinfo {eid} {arXiv:1807.05226}} (\bibinfo {year} {2018})},\ \Eprint
  {http://arxiv.org/abs/1807.05226} {arXiv:1807.05226 [astro-ph.HE]}
  \BibitemShut {NoStop}%
\bibitem [{\citenamefont {Schutz}(2011)}]{Schutz_2011}%
  \BibitemOpen
  \bibfield  {author} {\bibinfo {author} {\bibfnamefont {B.~F.}\ \bibnamefont
  {Schutz}},\ }\href {\doibase 10.1088/0264-9381/28/12/125023} {\bibfield
  {journal} {\bibinfo  {journal} {Class.~Quantum~Grav.}\ }\textbf {\bibinfo
  {volume} {28}},\ \bibinfo {pages} {125023} (\bibinfo {year}
  {2011})}\BibitemShut {NoStop}%
\bibitem [{\citenamefont {{Usman}}\ \emph {et~al.}(2019)\citenamefont
  {{Usman}}, \citenamefont {{Mills}},\ and\ \citenamefont
  {{Fairhurst}}}]{2019ApJ...877...82U}%
  \BibitemOpen
  \bibfield  {author} {\bibinfo {author} {\bibfnamefont {S.~A.}\ \bibnamefont
  {{Usman}}}, \bibinfo {author} {\bibfnamefont {J.~C.}\ \bibnamefont
  {{Mills}}}, \ and\ \bibinfo {author} {\bibfnamefont {S.}~\bibnamefont
  {{Fairhurst}}},\ }\href {\doibase 10.3847/1538-4357/ab0b3e} {\bibfield
  {journal} {\bibinfo  {journal} {\apj}\ }\textbf {\bibinfo {volume} {877}},\
  \bibinfo {eid} {82} (\bibinfo {year} {2019})},\ \Eprint
  {http://arxiv.org/abs/1809.10727} {arXiv:1809.10727 [gr-qc]} \BibitemShut
  {NoStop}%
\bibitem [{\citenamefont {{Aasi}}\ \emph {et~al.}(2015)\citenamefont {{Aasi}}
  \emph {et~al.}}]{2015CQGra..32g4001L}%
  \BibitemOpen
  \bibfield  {author} {\bibinfo {author} {\bibfnamefont {J.}~\bibnamefont
  {{Aasi}}} \emph {et~al.},\ }\href {\doibase 10.1088/0264-9381/32/7/074001}
  {\bibfield  {journal} {\bibinfo  {journal} {Class.~Quantum~Grav.}\ }\textbf
  {\bibinfo {volume} {32}},\ \bibinfo {eid} {074001} (\bibinfo {year}
  {2015})},\ \Eprint {http://arxiv.org/abs/1411.4547} {arXiv:1411.4547 [gr-qc]}
  \BibitemShut {NoStop}%
\bibitem [{\citenamefont {{Manzotti}}\ and\ \citenamefont
  {{Dietz}}(2012)}]{2012arXiv1202.4031M}%
  \BibitemOpen
  \bibfield  {author} {\bibinfo {author} {\bibfnamefont {A.}~\bibnamefont
  {{Manzotti}}}\ and\ \bibinfo {author} {\bibfnamefont {A.}~\bibnamefont
  {{Dietz}}},\ }\href@noop {} {\bibfield  {journal} {\bibinfo  {journal} {arXiv
  e-prints}\ } (\bibinfo {year} {2012})},\ \Eprint
  {http://arxiv.org/abs/1202.4031} {arXiv:1202.4031 [gr-qc]} \BibitemShut
  {NoStop}%
\bibitem [{\citenamefont {{LIGO Scientific Collaboration}}(2018)}]{lalsuite}%
  \BibitemOpen
  \bibfield  {author} {\bibinfo {author} {\bibnamefont {{LIGO Scientific
  Collaboration}}},\ }\href {\doibase 10.7935/GT1W-FZ16} {\enquote {\bibinfo
  {title} {{LIGO} {A}lgorithm {L}ibrary - {LALS}uite},}\ }\bibinfo
  {howpublished} {Free software (GPL)} (\bibinfo {year} {2018})\BibitemShut
  {NoStop}%
\bibitem [{\citenamefont {Hannam}\ \emph {et~al.}(2014)\citenamefont {Hannam},
  \citenamefont {Schmidt}, \citenamefont {Boh\'e}, \citenamefont {Haegel},
  \citenamefont {Husa}, \citenamefont {Ohme}, \citenamefont {Pratten},\ and\
  \citenamefont {P\"urrer}}]{PhysRevLett.113.151101}%
  \BibitemOpen
  \bibfield  {author} {\bibinfo {author} {\bibfnamefont {M.}~\bibnamefont
  {Hannam}}, \bibinfo {author} {\bibfnamefont {P.}~\bibnamefont {Schmidt}},
  \bibinfo {author} {\bibfnamefont {A.}~\bibnamefont {Boh\'e}}, \bibinfo
  {author} {\bibfnamefont {L.}~\bibnamefont {Haegel}}, \bibinfo {author}
  {\bibfnamefont {S.}~\bibnamefont {Husa}}, \bibinfo {author} {\bibfnamefont
  {F.}~\bibnamefont {Ohme}}, \bibinfo {author} {\bibfnamefont {G.}~\bibnamefont
  {Pratten}}, \ and\ \bibinfo {author} {\bibfnamefont {M.}~\bibnamefont
  {P\"urrer}},\ }\href {\doibase 10.1103/PhysRevLett.113.151101} {\bibfield
  {journal} {\bibinfo  {journal} {Phys.~Rev. Lett.}\ }\textbf {\bibinfo
  {volume} {113}},\ \bibinfo {pages} {151101} (\bibinfo {year}
  {2014})}\BibitemShut {NoStop}%
\bibitem [{\citenamefont {Schmidt}\ \emph {et~al.}(2015)\citenamefont
  {Schmidt}, \citenamefont {Ohme},\ and\ \citenamefont
  {Hannam}}]{PhysRevD.91.024043}%
  \BibitemOpen
  \bibfield  {author} {\bibinfo {author} {\bibfnamefont {P.}~\bibnamefont
  {Schmidt}}, \bibinfo {author} {\bibfnamefont {F.}~\bibnamefont {Ohme}}, \
  and\ \bibinfo {author} {\bibfnamefont {M.}~\bibnamefont {Hannam}},\ }\href
  {\doibase 10.1103/PhysRevD.91.024043} {\bibfield  {journal} {\bibinfo
  {journal} {Phys.~Rev.}\ }\textbf {\bibinfo {volume} {D91}},\ \bibinfo {pages}
  {024043} (\bibinfo {year} {2015})}\BibitemShut {NoStop}%
\bibitem [{\citenamefont {{Foreman-Mackey}}\ \emph {et~al.}(2013)\citenamefont
  {{Foreman-Mackey}}, \citenamefont {{Hogg}}, \citenamefont {{Lang}},\ and\
  \citenamefont {{Goodman}}}]{2013PASP..125..306F}%
  \BibitemOpen
  \bibfield  {author} {\bibinfo {author} {\bibfnamefont {D.}~\bibnamefont
  {{Foreman-Mackey}}}, \bibinfo {author} {\bibfnamefont {D.~W.}\ \bibnamefont
  {{Hogg}}}, \bibinfo {author} {\bibfnamefont {D.}~\bibnamefont {{Lang}}}, \
  and\ \bibinfo {author} {\bibfnamefont {J.}~\bibnamefont {{Goodman}}},\ }\href
  {\doibase 10.1086/670067} {\bibfield  {journal} {\bibinfo  {journal} {PASP}\
  }\textbf {\bibinfo {volume} {125}},\ \bibinfo {pages} {306} (\bibinfo {year}
  {2013})},\ \Eprint {http://arxiv.org/abs/1202.3665} {arXiv:1202.3665
  [astro-ph.IM]} \BibitemShut {NoStop}%
\bibitem [{\citenamefont {{Abbott}}\ \emph {et~al.}(2017)\citenamefont
  {{Abbott}} \emph {et~al.}}]{2017ApJ...848L..12A}%
  \BibitemOpen
  \bibfield  {author} {\bibinfo {author} {\bibfnamefont {B.~P.}\ \bibnamefont
  {{Abbott}}} \emph {et~al.},\ }\href {\doibase 10.3847/2041-8213/aa91c9}
  {\bibfield  {journal} {\bibinfo  {journal} {ApJL}\ }\textbf {\bibinfo
  {volume} {848}},\ \bibinfo {eid} {L12} (\bibinfo {year} {2017})},\ \Eprint
  {http://arxiv.org/abs/1710.05833} {arXiv:1710.05833 [astro-ph.HE]}
  \BibitemShut {NoStop}%
\bibitem [{\citenamefont {{Mortlock}}\ \emph {et~al.}(2018)\citenamefont
  {{Mortlock}}, \citenamefont {{Feeney}}, \citenamefont {{Peiris}},
  \citenamefont {{Williamson}},\ and\ \citenamefont
  {{Nissanke}}}]{2018arXiv181111723M}%
  \BibitemOpen
  \bibfield  {author} {\bibinfo {author} {\bibfnamefont {D.~J.}\ \bibnamefont
  {{Mortlock}}}, \bibinfo {author} {\bibfnamefont {S.~M.}\ \bibnamefont
  {{Feeney}}}, \bibinfo {author} {\bibfnamefont {H.~V.}\ \bibnamefont
  {{Peiris}}}, \bibinfo {author} {\bibfnamefont {A.~R.}\ \bibnamefont
  {{Williamson}}}, \ and\ \bibinfo {author} {\bibfnamefont {S.~M.}\
  \bibnamefont {{Nissanke}}},\ }\href@noop {} {\bibfield  {journal} {\bibinfo
  {journal} {arXiv e-prints}\ ,\ \bibinfo {eid} {arXiv:1811.11723}} (\bibinfo
  {year} {2018})},\ \Eprint {http://arxiv.org/abs/1811.11723} {arXiv:1811.11723
  [astro-ph.CO]} \BibitemShut {NoStop}%
\end{thebibliography}%

\end{document}